\documentclass[epj]{svjour} 
\usepackage{graphics} 
\usepackage{graphicx}
\usepackage{dcolumn}
\usepackage{amssymb} 
 
\usepackage{amsfonts} 
\usepackage{amssymb} 
\usepackage{amsmath} 
\usepackage{widetext}
\usepackage{amssymb} 
\usepackage{color}

\newcommand{\ar}{\arrowvert} 
\newcommand{\ra}{\rangle} 
 
\newcommand{\da}{\dagger} 
\newcommand{\ov}{\overline} 
\newcommand{\cd}{\! \cdot \!} 
\newcommand{\be}{\begin{equation}} 
\newcommand{\ee}{\end{equation}} 
\newcommand{\ba}{\begin{eqnarray}} 
\newcommand{\ea}{\end{eqnarray}}

\newcommand{\pa}{\partial} 

\begin{document} 
\title{Charm diffusion in a pion gas\\
implementing unitarity, chiral and heavy quark symmetries
}
\author{  
Luciano M. Abreu$^1$, Daniel Cabrera$ ^2$ , Felipe J. Llanes-Estrada$ ^3$  and Juan M. Torres-Rincon$ ^3$ 
}                     
 
\institute{ $^{1}$Instituto de F{\'i}sica, Universidade Federal da
Bahia, 40210-340, Salvador, BA, Brazil \\
$ ^2$ 
Departamento de F\'{\i}sica Te\'orica II, and \\
$^3$
Departamento de F\'{\i}sica Te\'orica I,  Universidad 
Complutense, 28040 Madrid, Spain. 
} 
\date{Received: date / Revised version: date} 
%
 
\abstract{  
We compute the charm drag and diffusion coefficients in a hot pion gas, such as is formed in a Heavy Ion Collision after the system cools sufficiently to transit into the hadron phase. We fully exploit Heavy Quark Effective Theory (with both $D$ and $D^*$ mesons as elementary degrees of freedom during the collision) and Chiral Perturbation Theory, and employ standard unitarization to reach higher temperatures. We find that a certain friction and shear diffusion coefficients are almost $p^2$-independent at fixed temperature which simplifies phenomenological analysis.\\
At the higher end of reliability of our calculation, $T\simeq 150$ MeV, we report a charm relaxation length $\lambda_c \simeq 40$ fm, in agreement with the model estimate of He, Fries and Rapp. \\
The momentum of a 1-GeV charm quark decreases about 50 MeV per Fermi when crossing the hadron phase.
} 
\authorrunning{Abreu, Cabrera, Llanes-Estrada and Torres-Rincon} 
\titlerunning{Charm diffusion in a pion gas...}
\maketitle

\section{Introduction} 

Heavy Ion Collisions provide a thriving branch of nuclear and particle physics. Thanks to technological advances in the last three decades, measurements that once looked too challenging can now be performed. One of these is the reconstruction of charmed and bottomed mesons flowing out of the nuclear debris, that modern vertex detectors, together with good particle reconstruction and the ability to automatically treat very large data samples have brought to the realm of measurability.

Heavy--flavored hadrons are interesting because the\\
 hadron medium is not hot enough to excite charm pairs. They are produced by hard gluons in the initial stages of the collision and their spectra will carry a memory of it, unlike pions and kaons that can be produced in the thermal medium at later stages, and thus show a spectrum close to black-body without much information from the initial configuration of fields. 

However, charmed and bottomed mesons do interact with the hadron gas after the crossover from the high--energy phase (that, although now known to be strongly coupled, we will continue naming ``quark--gluon plasma'' as is customary).  The corrections to their properties due to this cooler medium
requires their scattering cross--section with the medium pions and other particles. Given the scattering amplitudes one can proceed to kinetic simulations following individual particles, or employ kinetic theory to compute transport coefficients that can be input to bulk hydrodynamic simulations.

In this article we will be concerned with charmed mesons, of more immediate interest, although the theory developed can immediately be applied to bottomed mesons too, which we will leave for a future application.
 
The scattering amplitudes or cross sections for heavy mesons cannot be directly accessed by experiment (since the short life of these mesons makes impossible to focuse beams of them on a target) so their knowledge requires theory constraints.
In the past~\cite{Svetitsky:1996nj} cross--sections were only guessed on the basis of constituent quark counting. Since $\sigma_{pp}$ is about $40$ mbarn and $\sigma_{\psi N}$ about $2$ mbarn, this counting leads to a charm quark-light quark scattering cross-section which is $\sigma_{cq}\simeq 0.3$ mbarn, much smaller than $\sigma_{qq}\simeq 4$ mbarn, leading to $\sigma_{D\pi}\simeq 9 $ mbarn. 

As we will see, this old reasoning is not too much off the mark, but the cross-sections can now be accessed with more reliable theoretical methods~\cite{Fuchs,Lutz:2007sk,Guo:2009ct,Geng:2010vw,Gamermann:2007fi,Liu:2009uz,Tolos:2009nn}, combining Chiral Perturbation Theory, Heavy Quark Effective Theory, and Unitarity.

Given the renewed experimental interest, it appears that several theoretical groups have simultaneously been attempting to extract the transport coefficients from the increased understanding of hadron-hadron interactions.

\subsection{Current theoretical understanding and setup}

The work of Laine~\cite{Laine:2011is} employs canonical perturbation theory in HQET and ChPT and thus focuses on the lowest possible temperatures. Two simultaneous papers of He, Fries and Rapp~\cite{He:2011yi}
and of Ghosh {\it{et al.}}~\cite{Ghosh:2011bw} have attempted to reach higher temperatures, close to the cross-over to the quark and gluon plasma, by including further species of particles ($K$ and $\eta$ mesons or nucleons). While the second combines the perturbative approach of Laine with Born exchange terms, the first relies back on phenomenological estimates of the cross--sections.

We feel that there is still room for our contribution. There are serious disagreements among the three works cited. The extension of Ghosh {\it{et al.}} to higher energies does not include unitarity as a guiding principle, thus likely overestimating the cross-section since the polynomial perturbative expansion grows very fast with $s$.

By performing a state of the art computation of the pion-charmed hadron interactions, extending the work of Laine and Ghosh {\it{et al.}} by providing both the canonical HQET+ChPT perturbation theory \emph{and} unitarity, and tying the unknown parameters to experimental $D_0$ and $D_1$ resonances, we believe we have an interaction that is both solidly grounded in theory, and phenomenologically acceptable, drawing from the best features of the extant works.

As the charm transport coefficients are concerned, we will consider the drag or friction force $F$ (variously denoted $\gamma$, $\eta$ or $A$ in the literature), and the two $\Gamma_0$ and $\Gamma_1$ momentum-space diffusion coefficients. Other works have considered only isotropic drag and diffusion, in which case there is only one diffusion coefficient also denoted as $\kappa$ or $B_0$. We do not make this hypothesis and provide both coefficients corresponding to parallel and shear momentum transfers. Finally, in the $p\to 0$ limit, we make contact with the traditional kinetic theory and compute the space diffusion coefficient $D_x$ (again, sometimes denoted $D_s$ in the literature, but we avoid this notation to prevent confusion with the meson of equal name).  We find important to lift the hypothesis of isotropy because of the interesting elliptic flow observable.

Finally we make an additional contribution, in the philosophy of fully exploiting Heavy Quark Effective Theory (in addition to ChPT) as a starting point. In the Heavy Quark limit, the $D$ and $D^*$ mesons are degenerate and there are four (one spin-zero and three spin-one) propagating modes for the charm quark in the pion medium. This has been missed by all existing approaches, that only attend to $D$-propagation in-medium (Laine however considers both $B$ and $B^*$ in the bottom sector, where the $B^*$ meson is stable under strong decays). 

In the physical world the $D^*$ meson is unstable and decays to $D\pi$. However it does so with a small width (given its closeness to threshold) and thus, for the duration of the hadron gas, expected to be of the order of $5-10$ fm$/c$ at most, it propagates as a stable mode. To settle this point let us realize that, for a particle to decay within $5$ fm of its production point, its width has to be of order $40$ MeV, which, even after accounting for in-medium modifications (see Fig.~1 of~\cite{He:2011yi}), is only reached for $D^*$ mesons at temperatures of order the phase transition $T\simeq 180$ MeV, so that for the entire life of the hadron gas both $D$ and $D^*$ mesons need to be taken as elementary degrees of freedom.

We of course include the $DD^*\pi$ interaction vertex in the effective Lagrangian. However we will present computations in which the $D^*$ is thus included as an elementary particle (but also others without it for ease of comparison with the recent computations).  We will generically speak of the passage of the charm quark through the pion medium, whether hadronized in a $D$ or a $D^*$ meson. In the heavy quark limit, the heavy quark is little affected by the specific nature of the light degrees of freedom hadronizing around it.  

Given these theory improvements, and the fact that the other groups have not found very large effects from including strangeness or nucleons as explicit degrees of freedom in the hadron gas, we will comptent ourselves with examining the contribution of pions. A priori one can expect pions to provide the bulk of the charm-medium interaction, by their large multiplicity (typically one particle of any other species for every ten pions).

We employ the Fokker-Planck formalism for a heavy Brownian particle subject to the bombardment of the light pions in the medium. Our approximations will be sensible as long as the momentum of the heavy particle remains smaller than its mass in natural units, so that $p\ge 2$ GeV is not accessible by our computation (although we show plots at higher momentum for ease of comparison with future investigations addressing hard heavy flavors).
Pairs of heavy quarks rapidly drift apart and, by the time of the transition to the hadron phase, they are at least three Fermi away from each other and never rescatter in it (unless initially in a bound charmonium state). Since they are very scarce, we neglect the interactions between charm pairs formed in different points of the collision.

\subsection{Experimental motivation}

One could conceive hydrodynamic calculations of the quark-gluon plasma
that would result in fits of $F$, $\Gamma_0$ and $\Gamma_1$ to experimental data under certain assumptions on the initial distribution of heavy quarks. 
The information gained would be very valuable to understand how strongly that plasma is coupled, and perhaps restrict the possible initial state configurations.

However, the extraction of the coefficients is blurred by the hadron phase in the final state, as the system must cool before total freeze out, and charm quarks propagating through the resulting hadron medium will also suffer drag and diffusion. It is like trying to deduce the dispersive properties of a glass with a beam of light going through an additional lens: both have to be simultaneously understood.

Existing data on nuclear suppression factors and elliptic flow (see Sec.~\ref{sec:exp} below) have already been compared with standing calculations within the asymptotic quark-gluon plasma phase~\cite{Moore:2004tg} and, perhaps more successfully, with a mixed approach that includes resonances surviving into the plasma phase~\cite{vanHees:2005wb}.

Another observable that is being addressed in the literature is the transverse momentum spectrum of the $D$ mesons, that should be a rough thermometer of the phase transition~\cite{Svetitsky:1996nj}, provided that the effect of the final stage hadron phase does not blur all information out (it doesn't, as we will show in this article).

\subsection{The $D$-meson spectrum}

A charm quark propagating in the low-temperature medium below the deconfinement
phase transition must do so confined in a hadron. 
In central heavy-ion collisions the baryon number is very small and can be neglected.
Therefore one expects the charm quark to form a $D$-meson or an excitation thereof.
Let us briefly recall what experimental knowledge there is about the $D$-spectrum.

The ground state $D$-meson is as usual in meson spectroscopy a pseudoscalar $J^P=0^-$ 
with four charge states $+,-,0,\bar{0}$ (identified in the quark model as
$c\bar{d}$, $d\bar{c}$, $c\bar{u}$ and $u\bar{c}$ respectively in a relative
$s$-wave with spins antiparallel). 
Since we neglect isospin-breaking terms, we can average  the masses over this 
quartet to obtain $M_D\simeq 1867$ MeV. \\
This meson cannot decay by any strong process and we will take it to be absolutely
stable.

The first excitation is the vector $1^-$ $D^*$ meson whose mass average is
$M_{D^*}=2008.5$ MeV. In the Heavy Quark Limit this meson should degenerate with the
$D$, (and in fact this is seen by glancing higher to the $B$-meson whose splitting to the $B^*$ is much smaller).
This mass is barely above $D\pi$ threshold, so there is only this one strong decay
channel, and it is very suppressed.
 
The width of the charged $D^*$ is estimated at $1$ MeV, and that for the neutral 
partners has not been measured but is consistent with $\Gamma \le 2$ MeV.
This means that a $D^*$ has a mean lifetime in vacuum of order 100-200
fm. Since the typical freeze-out time of a heavy ion collision is about 20 fm
it is not a bad first approximation to take the $D^*$ meson as also stable during
the fireball's lifetime: there is ample room even in medium since the decay
time is an order of magnitude larger than the freeze-out time. As stated above, in-medium corrections do not alter the picture. Thus, most
$D^*$ mesons decay after collisions have ceased. This approximation can be 
corrected if wished by taking into account the in-medium inelastic process 
$D^*\to D \pi$ with Bose enhancement for the final-state pion.

\begin{figure}
\centering
\includegraphics[width=8.5cm]{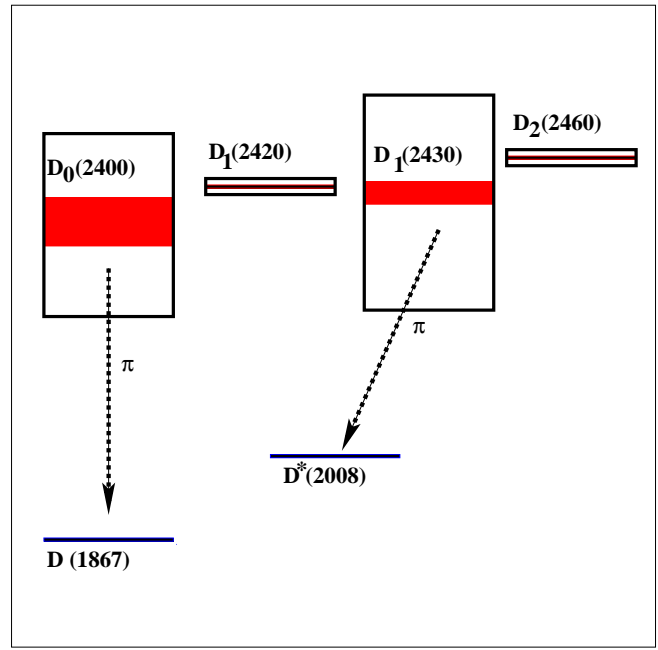}
\caption{\label{fig_spectrum} The currently known low-lying $D$-meson system.
The negative parity state $D$ and $D^*$ are represented as the blue lines.
The four positive parity states have the mass measurement spread throughout the
red boxes, while the hollow black boxes represent current estimates of their width.
$s$-wave pion decays are depicted.}
\end{figure}

In agreement with quark model expectations, the next-higher excitations of the
$D$ system seem to be a triplet and a singlet of positive parity, with spins
$0^+,1^+,2^+$ and $1^+$ respectively, corresponding to $ ^{2S+1}L_J= 
^3P_J$ and $ ^1P_1$. 
The two mesons with spin 1 and positive parity must mix, and they do so in an 
interesting manner: the one with lowest mass, $D_1(2420)$ becomes narrow and hence
decoupled from the natural $s$-wave decay channel $D^*\pi$, whereas the 
higher member $D_1(2430)$ is very broad and seen in that configuration.
The situation can be seen in Fig.~\ref{fig_spectrum} and in Table~\ref{table_spectrum}.

\begin{table}[h]
\caption{Charged-average masses and experimental estimates~\cite{Nakamura:2010zzi} for the strong 
widths of the $D$-meson resonances. Units are MeV. Errors not quoted are
about $1$ MeV or less. \label{table_spectrum}}
\centering
\begin{tabular}{|c c||c c|}
\hline 
Meson & $J^P$ & $M$ (MeV) & $\Gamma$ (MeV) \\
\hline
$D $  & $0^-$ & 1867     & -       \\
$D^*$ & $1^-$ & 2008     & 1       \\
$D_0$ & $0^+$ & 2360(40) & 270(50) \\
$D_1$ & $1^+$ & 2422     & 22(5)   \\
$D_1$ & $1^+$ & 2427(40) & 380(150)\\
$D_2$ & $2^+$ & 2460     & 30      \\
\hline
\end{tabular}
\end{table}

The remaining low-lying resonance, the $D_2$, is again narrow. Since its mass at 
2460 MeV is 600 MeV above the ground-state $D$ meson, and it is quite decoupled
due to its moderate width of about $40$ MeV, we do not expect this (nor the $D_1(2420)$
to play an important role at small temperatures.

Thus a sensible approach to charm propagation in a heavy-ion collision after the
phase transition to a hadron gas has occurred, is to take the $D$ and $D^*$ 
mesons as absolutely stable degrees of freedom for the $c$-quark, that in 
collision with the in-medium pions they rescatter into the resonances $D_0$ and
$D_1(2430)$.

The experimental knowledge of the resonances $D_0$ and $D_1$ sufficiently constrains the low-energy effective Lagrangian density for $c\pi$ scattering so that we are in possession of a good approximation to the cross section.

\section{The Fokker-Planck equation} 
\subsection{Derivation}
The momentum-space distribution of charm quarks with momentum $p$, $f_c(p)$, is not in equilibrium when the hadron phase of a heavy-ion collision forms, and must relax via a Boltzmann equation.
\be \frac{df_c(\mathbf{p})}{dt} = C [f_c(\mathbf{p})] \ , \ee
The right hand side is called the collision operator, because
it describes kinetic collisions of the charmed particles. 
The left hand side, in the absence of external forces, is the advective derivative
\be \frac{\partial f_c(\mathbf{p})}{\pa t} + \mathbf{v} \cdot {\vec{\nabla}}_x f_c(\mathbf{p}) = 
\left[ \frac{\pa f_c(\mathbf{p})}{\pa t} \right]_{coll} \ .\ee

The density of $D$ and $D^*$ mesons being very small, 
we can neglect collisions between $D$ mesons themselves and concentrate only on the interaction of these charmed mesons with the pion bath, assumed in thermal equilibrium.

The bath's distribution function $f_{\pi} (\mathbf{q})$  is hence the Bose-Einstein function. Moreover, the gas is assumed homogeneous and the distribution does not depend on $\mathbf{x}$. For this reason one can average the Boltzmann equation over the collision volume and understand the 1-particle distribution function for the charmed mesons as
the average
\be f_c (t, \mathbf{p}) \equiv \frac{1}{V} \int d \mathbf{x} \ f_c(t,\mathbf{x}, \mathbf{p}) \ .\ee
The averaged Boltzmann equation becomes then
\be \frac{\pa f_c (t , \mathbf{p})}{\pa t} = \left[ \frac{\pa f_c (t , \mathbf{p})}{\pa t} \right]_{coll} \ .\ee

Charmed mesons may enter and exit the momentum element $d{\bf p}$ around ${\bf p}$ by collisions with the pion bath, so the collision term has two parts associated with gains and losses.

Gains in the momentum distribution around ${\bf p}$ are proportional to the probability density around $(\mathbf{p}+\mathbf{k})$ times the probability of transferring momentum $\mathbf{k}$ from the charmed meson to the bath.
It is therefore convenient to define a collision rate $w(\mathbf{p},\mathbf{k})$ for a charmed meson with initial and final momenta $\mathbf{p}$, $\mathbf{p}-\mathbf{k}$.

Conversely, losses are proportional to the distribution function around 
$\mathbf{p}$ times the probability of transferring momentum $\mathbf{k}$ to the pion bath.

In principle, 
the Boltzmann equation should be treated as a quantum
Boltzmann-Uehling-Uhlenbeck equation taking into account Bose enhancement effect in the final state, with factors $(1+f_c)$ that encode 
the increased probability of a charmed meson scattering into an already occupied state, 
\ba
\frac{\pa f_c (t , \mathbf{p})}{\pa t} = 
\\ \nonumber 
\int d\mathbf{k} \left\{ f_c(t,\mathbf{p}+\mathbf{k}) w (\mathbf{p}+ \mathbf{k}, \mathbf{k}) 
\left[1+f_c (t,\mathbf{p}) \right] \right.
\\ \nonumber  \left.
-  f_c(t,\mathbf{p}) w (\mathbf{p}, \mathbf{k} ) \left[ 1+f_c (t, \mathbf{p}-\mathbf{k}) \right] \right\}
\ . 
\ea

However, as the number of $c$-quarks is very small, we can approximate 
$1+f_c(t, \mathbf{p}) \approx 1$ inside the collision operator in practice.  This approximation however is probably not valid for the pion distribution function and we keep the $(1+f_\pi)$ factor in Eq.~(\ref{probdist}) below.  
As the charmed mesons are concerned, a classical Boltzmann equation should however be very accurate,
\be 
\frac{\pa f_c (t , \mathbf{p})}{\pa t} = \int d\mathbf{k} \left[ f_c(t,\mathbf{p}+\mathbf{k}) w (\mathbf{p}+ \mathbf{k}, \mathbf{k}) -  f_c(t,\mathbf{p}) w (\mathbf{p}, \mathbf{k} ) \right]
\ .\ee
In turn the collision rate  can be spelled out in terms of the Lorentz invariant charm quark-pion scattering amplitude,

\begin{widetext}
\be \label{probdist}
w (\mathbf{p}, \mathbf{k} ) = g_{\pi} \int \frac{d \mathbf{q}}{(2\pi)^9} f_{\pi} (\mathbf{q}) \left[ 1+ f_{\pi} (\mathbf{q}+\mathbf{k}) \right]
 \frac{1}{2E_q^{\pi}} \frac{1}{2E_p^{c}} \frac{1}{2 E_{q+k}^{\pi}} \frac{1}{2 E_{p-k}^c}
 (2\pi)^4 \delta (E_p^c + E_q^{\pi} - E^c_{p-k} -E_{q+k}^{\pi} )  \sum |\mathcal{M}_{\pi c}(s,t,\chi)|^2
\ee
\end{widetext}
\noindent
($g_\pi=3$ is the pion isospin degeneracy, and $\chi$ denotes the possible spin degrees of freedom, active if the $c$ quark finds itself inside a $D^*$ meson).  The scattering amplitude $\mathcal{M}$ is normalized according to standard covariant convention~\cite{Nakamura:2010zzi}. Note that Eq.~(7) of~ \cite{He:2011yi} differs by the Bose-enhancement factor $(1+f_\pi)$ for the pion exiting the collision. We believe that in the temperature range of $m_\pi\simeq T \simeq 150$ MeV that we (and those authors) treat, this enhancement should not be neglected.

The Boltzmann equation in this case reduces to a much simpler Fokker-Planck equation  because the mass of the $D$ and $D^*$ mesons carrying the $c$-quark is much greater than the mass of the pions and the temperature of the heat bath.
Then, the scale of momentum for which there is a significant change of $f_c(p)$ with the momentum of the $D$ meson $|\mathbf{p}|$ is greater than the typical transfered momentum $|\mathbf{k}|$, that is of the order of $T$:
\be |\mathbf{p}|_{\rm{f_c}} \gg |\mathbf{k}| \sim T \sim 150 \textrm{ MeV} \ . \ee

Because of this separation of scales, it is natural to expand the collision rate inside the collision operator respect to its first argument ${\bf p}+{\bf k}$,
\ba \label{colintegral}
wf \equiv w(\mathbf{p}+\mathbf{k},\mathbf{k}) \ f_c(t, \mathbf{p} + \mathbf{k}) = \\ \nonumber
w (\mathbf{p},\mathbf{k}) f_c (t,\mathbf{p}) + k_i \frac{\pa}{\pa p_i} (wf) + \frac{1}{2} k_i k_j \frac{\pa ^2}{\pa p_i \pa p_j} (wf) \dots
\ea
with $i,j=1,2,3$.
The collision integral reads, with this substitution,
\be \left[ \frac{\pa f_c (t , \mathbf{p})}{\pa t} \right]_{coll} = \int d\mathbf{k} \left[ k_i \frac{\pa}{\pa p_i} + \frac{1}{2} k_i k_j \frac{\pa ^2}{\pa p_i \pa p_j} \right] \left( wf \right)\ .
\ee
This suggests defining two auxiliary functions, 
\begin{eqnarray}
 F_i (\mathbf{p}) & = & \int d\mathbf{k} \ w(\mathbf{p},\mathbf{k}) \ k_i \ ,\\
\Gamma_{ij} (\mathbf{p}) & = & \frac{1}{2} \int d\mathbf{k} \ w(\mathbf{p},\mathbf{k}) \ k_i k_j\ , 
\end{eqnarray}
whose classical interpretation is that of a drag force acting on the charmed particle, and the autocorrelation of a random, Brownian force, as will be shown below in Appendix~\ref{langevin}.

Eq.~(\ref{colintegral}) reduces to the Fokker-Planck equation
\be \label{FKPL}
\frac{\pa f_c (t, \mathbf{p})}{\pa t} = \frac{\pa}{\pa p_i} \left\{ F_i (\mathbf{p}) f_c (t,\mathbf{p}) + \frac{\pa}{\pa p_j} \left[ \Gamma_{ij} (\mathbf{p})  f_c (t,\mathbf{p}) \right] \right\}
\ee
where meanwhile we can see that $F_i$ behaves as a friction term representing the average momentum change of the $D$ meson and $\Gamma_{ij}$ acts as a diffusion coefficient in momentum space, as it forces a broadening of the average momentum distribution of the $D$ meson. This interpretation also falls-off from the one-dimensional solution that we leave for Appendix~\ref{OnedFK}. 

We will not find necesary to solve the three-dimensional Fokker-Planck equation for $f_c(p)$ in full, but only to calculate the coefficients $F_i$ and $\Gamma_{ij}$ that already encode the physics of charm drag and diffusion.

\subsection{$F_i$ and $\Gamma_{ij}$ Coefficients}

In the ideal case where the pion gas is homogeneous and isotropic, and because the coefficients $F_i$ and $\Gamma_{ij}$ only depend on $p^i$, they can be expressed as a function of three scalar functions by means of
\begin{eqnarray}\label{defAyB}
F_i (\mathbf{p}) & = & F(p^2) p_i, \\ \nonumber
\Gamma_{ij} (\mathbf{p}) & = & \Gamma_0 (p^2) \Delta_{ij} + \Gamma_1 (p^2) \frac{p_i p_j}{p^2} \ ,
\end{eqnarray}
where
\be \Delta_{ij} \equiv \delta_{ij} - \frac{p_i p_j}{p^2}
\ee
satisfies the handy identity $\Delta_{ij} \Delta^{ij} =2$.

%

We choose the momenta of the elastic collision between a charmed meson $D$ or $D^*$ and a pion as
\be D (\mathbf{p}) + \pi (\mathbf{q}) \rightarrow D(\mathbf{p}-\mathbf{k}) + \pi (\mathbf{q} +\mathbf{k}). \ee

The three scalar coefficients in Eq.~(\ref{defAyB}) are then simple integrals over the interaction rate
\begin{eqnarray} \label{Transportintegrals}
 F(p^2) & = & \frac{p^i F_i}{p^2} = \int d\mathbf{k}\  w(\mathbf{p},\mathbf{k})  \ \frac{k_ip^i}{p^2} \ , \\ \nonumber
 \Gamma_0 (p^2) & = & \frac{1}{2} \Delta_{ij} \Gamma^{ij} = \frac{1}{4} \int d\mathbf{k}\ w(\mathbf{p},\mathbf{k}) \left[ \mathbf{k}^2 - \frac{(k_i p^i)^2}{p^2} \right] \ , \\ \nonumber
 \Gamma_1(p^2) & = & \frac{p_i p_j}{p^2} \Gamma^{ij} = \frac{1}{2} \int d\mathbf{k}\  w(\mathbf{p},\mathbf{k}) \ \frac{(k_i p^i)^2}{p^2} \ , 
\end{eqnarray}
where the dynamics is fed-in by the scattering matrix elements $|\mathcal{M}_{\pi c}|$.
The choice of kinematic integration variables and the reduction of these integrals is detailed in Appendix~\ref{Kinematicsapp}.

We also remind the reader in Appendix~\ref{langevin} how the interpretation of the friction coefficient times the quark momentum $F\ p$ is that of an energy loss per unit length upon propagation of the charm quark in the plasma, and how the loss of momentum per unit length is simply $F\ E$ in terms of energy and momentum of the charmed particle.

After we numerically control the cross-section and scattering amplitude $\mathcal{M}$ for the charm quark in the pion medium, we evaluate the three transport coefficients 
and give the results in subsection~\ref{subsec:numtransport} below.
We quote there two different approximations. One in which the $D^*$ is neglected as a propagating degree of freedom (akin to what can be found so far in the literature), and one in which the $c$-quark can travel also as a $D^*$ meson (with slightly modified interaction and kinematics).

\section{Effective Lagrangian for $D$,  $D^*$ and
$\pi$ with ChPT and HQET} 

Now we construct the chiral Lagrangian density that describes the interactions between the spin-0 and spin-1 $D$-mesons and pseudoscalar Goldstone bosons. The leading order (LO) chiral Lagrangian $\mathcal{L}^{(1)}$ is given by \cite{Lutz:2007sk,Guo:2009ct,Geng:2010vw},   
 \ba
\mathcal{L}^{(1)} & = &  \nabla ^{\mu} D \, \nabla _{\mu}D^{\da} - m_D ^2 D D^{\da} -  \nabla ^{\mu} D^{\ast \nu} \,\nabla _{\mu} D^{\ast \da}_{\nu} \nonumber \\
& &  + m_{ D} ^2 D^{\ast \mu}  D^{\ast \da}_{\mu} + ig \left(  D^{\ast \mu} u_{\mu}  D^{ \da} - D u^{\mu} D^{\ast
\da}_{\mu} \right) \nonumber \\
& & + \frac{g}{2 m_D} \left(  D^{\ast}_{\mu} u_{\alpha}   \nabla_{\beta} D^{\ast \da}_{\nu} -  \nabla_{\beta} D^{\ast}_{\mu} u_{\alpha}  D^{\ast \da}_{\nu} \right) \varepsilon ^{\mu \nu \alpha \beta} \ ,  \nonumber \\
\label{lag1}
\ea
where $D=(D^0,D^{+}, D^{+}_{s})$ and  $D^{\ast} _{\mu}=(D^{\ast 0}, D^{\ast + },D^{\ast +}_{s})_{\mu}$ are the SU(3) anti-triplets of spin-zero and spin-one $D$-mesons with the chiral limit mass $m_D$, respectively. We have also used the quantities
\ba
\nabla  _{\mu}  & = & \partial _{\mu} - \Gamma _{\mu}, \nonumber \\
\Gamma _{\mu} & = & \frac{1}{2} \left( u^{\da} \partial _{\mu} u +  u\partial _{\mu} u^{\da} \right), \nonumber \\
u _{\mu} & = & i \left( u^{\da} \partial _{\mu} u -  u\partial _{\mu} u^{\da} \right),
\label{cov_der}
\ea
where
\be
u = \sqrt{U} = \exp{\left( \frac{i \Phi}{\sqrt{2} F} \right)}
\label{u_field}
\ee
is the unitary matrix incorporating the pseudoscalar Goldstone bosons,   
\be
\Phi = \left( \begin{array}{ccc}
  \frac{1}{\sqrt{2}}\pi ^{0}+\frac{1}{\sqrt{6}} \eta  & \pi ^{+} & K^{+} \\
    \pi^{-} & -\frac{1}{\sqrt{2}}\pi ^{0}+\frac{1}{\sqrt{6}} \eta & K^0 \\
    K^{-}   & \bar{K}^{0}  &  -\frac{2}{\sqrt{6}} \eta
\end{array}
            \right).
\label{phi}
\ee
$F$ in Eq. (\ref{u_field}) is the Goldstone boson decay constant in chiral limit.

The NLO chiral Lagrangian $\mathcal{L}^{(2)}$ reads
 \ba
\mathcal{L}^{(2)} & = & - h_0  D D^{\da} <\chi_+> + h_1  D \chi_+ D^{\da} + h_2  D D^{\da} <u^{\mu}u_{\mu}> \nonumber \\
& & + h_3  D u^{\mu}u_{\mu} D^{\da} + h_4  \nabla _{\mu} D  \,  \nabla _{\nu} D^{\da}  <u^{\mu}u^{\nu}> \nonumber \\
& & + h_5  \nabla _{\mu} D   \{ u^{\mu}, u^{\nu} \} \nabla _{\nu}D^{\da} + \tilde{h}_0  D^{\ast \mu} D^{\ast \da}_{\mu} <\chi_+> \nonumber \\
& & - \tilde{h}_1  D^{\ast \mu} \chi_+ D^{\ast \da}_{\mu}  - \tilde{h}_2  D^{\ast \mu}  D^{\ast \da}_{\mu}  <u^{\nu} u_{\nu}> \nonumber \\
& &  - \tilde{h}_3  D^{\ast \mu}  u^{\nu}u_{\nu} D^{\ast \da}_{\mu} - \tilde{h}_4  \nabla _{\mu}D ^{\ast \alpha}  \,  \nabla _{\nu}D^{\ast \da}_{\alpha}  <u^{\mu}u^{\nu}> \nonumber \\
& & - \tilde{h}_5 \nabla _{\mu}D ^{\ast \alpha}  \{ u^{\mu}, u^{\nu} \}\nabla _{\nu}D^{\ast \da} _{\alpha} \ . 
\label{lag2}
\ea
where
\be
\chi _{+} =  u^{\da} \chi u^{\da} +u \chi u \ ,
\label{u_chi}
\ee
with $\chi = \mathrm{diag}(m^2_{\pi}, m^2_{\pi}, 2 m^2_{K} -m^2_{\pi})$ being the mass matrix. The twelve parameters $h_i, \tilde{h}_i (i=0,...,5)$ are the low-energy constants (LECs), to be determined. However, we can make use of some constraints to reduce the set of free LECs. First, it should be noticed that in the limit of large number of colors ($N_c$) of QCD~\cite{'tHooft:1973jz}, single-flavor trace interactions are dominant. So, we fix $h_0 = h_2= h_4 = \tilde{h}_0 = \tilde{h}_2 = \tilde{h}_4 = 0$ henceforth. Besides, by imposing the heavy-quark symmetry (as will become clear in subsection~\ref{subsec_hqs} below), it follows that $\tilde{h}_i \simeq  h_i $.

In the following, the lowest order of the perturbative expansion of the quantities $\Gamma _{\mu}$, $u _{\mu}$ and $\chi _{+} $ in Eqs. (\ref{lag1}) and (\ref{lag2}) is considered to construct the scattering matrix of the interactions between the charmed mesons and the pseudoscalar Goldstone bosons.

\section{Scattering matrix for the $c$ quark
in the pion gas}

\begin{figure}
\centering
\includegraphics[width=8.5cm]{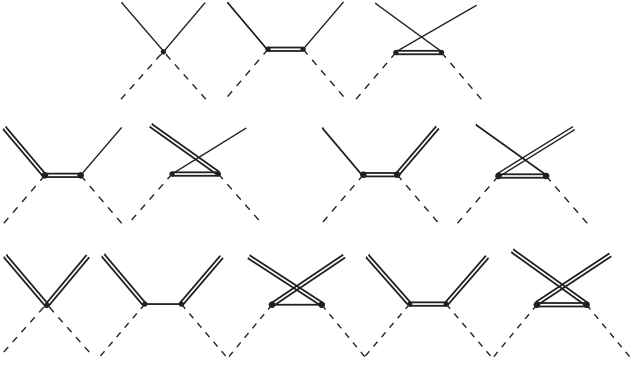}
\caption{Tree-level diagrams relevant to the scattering amplitudes in Eq. (\ref{ampl1}). The solid, double and dashed lines represent the $D$-mesons,
 $D^{\ast}$-mesons and Goldstone bosons, respectively.}
\label{tree}
\end{figure}

From the Lagrangian in Eq.~(\ref{lag1}) we are able to obtain the scattering amplitudes $V$ for $D, D^{\ast} \Phi \rightarrow D, D^{\ast} \Phi $ processes. In Fig. \ref{tree} we show the tree-level diagrams constructed from the LO and NLO interactions. These include both contact interactions and Born exchanges. 
The different scattering channels are labeled as
$V_{a}$ through $V_{d}$,
where the subscripts refer to the scattering channels as follows 
\ba
&(a)&:  D \phi \rightarrow D \phi, \nonumber \\
&(b)&:  D ^{\ast}\phi \rightarrow D \phi \nonumber \\
&(c)&:  D \phi \rightarrow D ^{\ast}\phi \nonumber \\
&(d)&:  D ^{\ast} \phi  \rightarrow D^{\ast} \phi \ .  
\label{proc}
\ea

Notice that, because of the scarcity of strange quarks in the heavy ion collison debris (kaon multiplicity is 10\% of typical pion multiplicity) we are interested only in channels involving the scattering between charmed mesons and pions with total strangeness equal to zero. So then we can simplify $\phi\to \pi$ and write down the relevant amplitudes as 

\begin{widetext}
\ba
V_{a} & = &  \frac{C_{0}}{4 F^2} (s - u)  + \frac{2C_{1}\,m_{\pi }^2}{ F^2} h_1   +  \frac{ 2 C_{2}}{ F^2} h_3 (p_2 \cdot p_4 ) +  \frac{2 C_{3}}{ F^2} h_5 \left[ (p_1 \cdot p_2 ) (p_3 \cdot p_4 ) + (p_1 \cdot p_4 )(p_2 \cdot p_3 ) \right] 
  \nonumber \\
& & + \frac{2 i\,g^2}{ F^2}  p_2^{\mu} \left[   C_{4}\,D_{\mu \nu } (p_1 + p_2)+  C_{5}\, D_{\mu \nu } (p_2 - p_3) \right] p_4 ^{\nu} \ ,\nonumber \\
V_{b} & = & \frac{ i \,g^2}{ m_D F^2} \left[  C_{4} \, p_2^{\alpha } \left( 2 p_1^{\beta } + p_2  ^{\beta }\right) p_{4\rho} 
D^{\nu \rho}(p_1 + p_2)   +  C_{5}\, p_4^{\alpha } \left( p_2^{\beta } - p_3  ^{\beta } -p_1^{\beta } \right) p_{2\rho} 
  D^{\nu \rho}(p_2 - p_3)  \right] \varepsilon _{\alpha \beta \mu \nu} \mathbf{\epsilon }^{\mu}(p_1) \ , \nonumber \\
 V_{c} & = & \frac{ i \,g^2}{ m_D F^2} \left[  C_{4}\, p_4^{\alpha } \left( p_1^{\beta } + p_2^{\beta } + p_3  ^{\beta }  \right) 
p_{2\rho}    D^{ \rho \nu }(p_1 + p_2)  +  C_{5} \, p_2^{\alpha } \left( p_2^{\beta } -2  p_3  ^{\beta } \right) p_{4\rho}  
D^{\nu \rho}(p_2 - p_3)  \right] \varepsilon _{\alpha \beta \mu \nu} \mathbf{\epsilon } ^{\ast \mu}(p_3) \ , \nonumber \\
V_{d} & = &  -\left\{ \frac{C_{0}}{4 F^2} (s - u)  + \frac{2C_{1}\,m_{\pi }^2}{ F^2} \tilde{h}_1   +  
\frac{ 2 C_{2}}{ F^2}\tilde{h}_3 (p_2 \cdot p_4) +  \frac{2 C_{3}}{ F^2} \tilde{h}_5 \left[ (p_1 \cdot p_2 ) (p_3 \cdot p_4 ) 
+ (p_1 \cdot p_4 )(p_2 \cdot p_3 ) \right] \right\} \epsilon ^{\mu}(p_1)\epsilon^{\ast} _{ \mu}(p_3)\nonumber \\ 
& & + \frac{2 i \,g^2}{ F^2}  \left[ C_{4}\, D (p_1 + p_2) +  C_{5} \, D(p_2 - p_3) \right] p_2^{\mu}\epsilon _{\mu}(p_1)  p_4 ^{\nu}
\epsilon^{\ast} _{ \nu}(p_3) \nonumber \\
& & +  \frac{ i g^2}{3 m_D^2 F^2} \left[  C_{6} \, p_2^{\alpha } \left( 2 p_1^{\beta } 
+ p_2  ^{\beta }\right) p_{4} ^{\rho} \left( p_1^{\sigma } + p_2  ^{\sigma } +  p_3  ^{\sigma }\right)  
D^{\nu \gamma}(p_1 + p_2) +  C_{7} \,  p_2^{\alpha } \left(  p_2^{\beta } - 2 p_3  ^{\beta }\right) p_{4} ^{\rho} \left( p_2^{\sigma }
 - p_3  ^{\sigma } -  p_1  ^{\sigma }\right) D^{\nu \gamma}(p_2 - p_3)   \right]\nonumber \\
& & \times \varepsilon _{\alpha \beta \mu \nu}\varepsilon _{\rho \sigma \gamma \delta} \mathbf{\epsilon } ^{\mu}(p_1) \mathbf{\epsilon }^{\ast \delta}(p_3) \ ,
\label{ampl1}
\ea  
\end{widetext}
\noindent
where $ C_{i}\;(i=0,...,7)$ are the coefficients of the scattering amplitudes for $D\pi ,D^{\ast} \pi$  channels with total isospin $I$, done in Table \ref{table2}, and $D (p)$, $D_{\mu \nu } (p)$ are the propagators of $D$ and $D^{\ast}$-mesons, respectively, 
\ba
D (p) & = & \frac{i}{p^{2} - m_D^2} \ , \nonumber \\
D^{\mu \nu }(p) & = & \frac{-i}{p^{2} - m_D^{*2}} \left( \eta^{\mu \nu} - \frac{p^{\mu}p^{\nu}}{m_D^{*2}}\right)\ .
\label{propagators}
\ea
As the two particles in all amplitudes are distinguishable, there is no $t$-channel type contribution (as e.g. in Compton scattering) with our relevant fields (open charm mesons and pions), and only $s$ and $u$-channel interactions appear. Between a $D$ and a $\pi$ one could exchange additional, closed flavor resonances in the $t$-channel, but a quick examination makes clear that these contributions are totally negligible. For example, $f_0$ exchange, while having strong coupling to two pions, has negligible coupling to two $D$ mesons, so one of the vertices makes the amplitude very small. Similarly, $J/\psi$ $t$-channel exchange is suppresed because of the small two-pion coupling of the very narrow state (and similar for other, closed flavor resonances). It doesn't make sense to include these resonances while neglecting higher order chiral and heavy quark corrections to the $D\pi$ Lagrangian with the basic fields.

Finally $\mathbf{\epsilon } ^{\mu}(p)$ is the polarization vector of the vector $D^{\ast}$-meson. If we were to write the polarization indices explicitly, 
 $\mathbf{\epsilon } ^{\mu}(p)\equiv  \mathbf{\epsilon}^{\mu}_\lambda(p)$,  
$V_b\equiv V_{b\lambda}$, $V_c\equiv V_{c\lambda}$, $V_d\equiv V_{d\lambda \lambda'}$, while $V_a$ remains a scalar as no vector mesons appear.

The amplitudes $V_b$ and $V_c$ must be related by time reversal, since they encode $D^*\pi \to D\pi$ and $D\pi\to D^*\pi$ respectively. Indeed, if one exchanges $p_1$ by $p_3$ and $p_2$ by $p_4$, and employs energy-momentum conservation $p_1+p_2=p_3+p_4$, they map onto each other as $V_b \to V_c$, $V_c\to V_b$.

\begin{table}[h]
\caption{Coefficients of the scattering amplitudes for the $D\pi ,D^{\ast} \pi$ channels with total isospin $I$ in Eq. (\ref{ampl1}).  \label{table2}}
\centering
\begin{tabular}{|c ||c c|}
\hline 
Constants &  $I= \frac{1}{2}$ & $I=\frac{3}{2} $   \\
\hline
$C_0 $  & $-2$ & 1       \\
$C_1 $  & $-1$ & $-1$    \\
$C_2 $  & $1$ & $1$       \\
$C_3 $  & $1$ & $1$       \\
$C_4 $  & $3$ & $0$       \\
$C_5 $  & $\frac{1}{3}$ & $\frac{2}{3}$       \\
$C_6 $  & $3$ & $0$       \\
$C_7 $  & $\frac{1}{3}$ & $\frac{2}{3}$ \\
\hline
\end{tabular}

\end{table}

\subsection{Heavy Quark Symmetry}\label{subsec_hqs}

Manifesting the spin of the vector meson, but ignoring the isospin index, Heavy Quark Symmetry should manifest itself~\cite{Laine:2011is} by the transformation 
\ba
\delta D = -\vec{\alpha}\cdot {\bf D}^* \\
\delta {\bf D}^* = \vec{\alpha} D + \vec{\alpha}\times {\bf D}^* \ . 
\ea
The Lagrangian density in Eq.~(\ref{lag2}) has been constructed manifestly maintaining chiral symmetry, that is then broken only carefully in perturbation theory upon expanding in fields and derivatives to construct the $LO$ and $NLO$ chiral amplitudes.

However since the charmed quarks are heavy fields, one should recover the Heavy Quark Symmetry by taking $m_D\to \infty$.

Referring to our amplitude in Eq.~(\ref{ampl1}), both $V_b$ and $V_c$ are of order $1/m_D$. To see it, one needs to write the denominator of the propagator as 
$$
(p^2-m_D^2)^{-1}\simeq ((p^0-m_D)\times 2m_D)^{-1}
$$ 
and notice that in the numerator at most one of the momenta can take the value $p^0\simeq m_D$ because of the antisymmetric Levi-Civita tensor: the other three four-vectors have to take spatial values to avoid its vanishing. Therefore the term within brackets is of $\mathcal{O}(1)$ in the $1/m_D$ counting and the explicit factor of
$1/m_D$ in front of the bracket suppresses the term.

Thus, the spin-changing amplitudes $D^*\pi \to D\pi$ and $D\pi\to D^*\pi$ are of $\mathcal{O}(1/m_D)$ and vanish in the heavy quark limit as expected, a collision with a pion cannot change the heavy-quark spin that decouples.

Turning to the elastic $D^*\pi\to D^*\pi$ amplitude, we notice that the last bracket of $V_d$ (carrying terms proportional to $C_6$ and $C_7$) is also $NLO$ in the heavy quark counting. We see that because of the two Levi-Civita $\varepsilon$ tensors, only one of the pair of $\alpha$, $\beta$ indices and only one of the pair $\rho$, $\sigma$ can take the value $0$. Because of the explicit propagator of order $1/m_D$, the bracket is at most of order $m_D$, and the $1/m_D^2$ factor in front of it suppresses it. 

The remaining part of the $V_d$ amplitude is then equal to $V_a$ if we impose $\tilde{h_i}=h_i$ as we have, yielding the expression of Heavy Quark Symmetry
$$
V_a(D\pi \to D\pi) = -V_d(D^*\pi\to D^*\pi)\ ,
$$
(the polarization of the $D^*$ will be handed shortly).
That is, in the infinitely heavy quark limit, the charmed mesons propagate in four states (the $D$ and the three polarizations of the vector $D^*$) that do not mix with each other and have equal scattering rates with the pion gas.

In this limit, the masses $m_D=m_D^*$ and also the dynamical resonances accessible in the scattering have equal mass $m_{D_0}=m_{D_1}$ and width
$\Gamma_{D_0}=\Gamma_{D_1}$.

Further, the Born exchange piece (terms proportional to $C_4$ and $C_5$) in both $D\pi$ and $D^*\pi$ elastic amplitudes is subleading in HQET. To demonstrate it, we expand the intermediate meson propagators
\ba
iD_{\mu\nu}(p_1+p_2) = \\ \nonumber
\frac{1}{(p_1+p_2)^2-m_D^2}
\left(\eta^{\mu\nu}-\frac{(p_1+p_2)^\mu(p_1+p_2)^\nu}{m_D^2} \right) \simeq 
\\ \nonumber
\frac{1}{(p_1^0)^2-m_D^2+2p_1^0 p_2^0 \dots} 
\left( \eta^{\mu\nu}-\delta^{\mu 0}\delta^{\nu 0}\frac{p_1^0p_1^0}{m_D^2} \right) \simeq
\\ \nonumber
\frac{1}{2m_D E^{\pi}_{2}} \left( \eta^{\mu\nu}- \delta^{\mu 0}\delta^{\nu 0}
\right) \ ,
\ea
suppressed by $m_D^{-1}$ as claimed.

The sum over polarizations also simplifies in the heavy quark limit. 
The vector-meson polarization basis then becomes a conventional spacelike spin-1 basis tied to a fixed reference frame, in the Cartesian basis simply
\be
\epsilon^\mu(p,\lambda)\simeq \epsilon^\mu(\lambda)=(0,\hat{\bf e}_\lambda)
\ee
satisfying the closure relation ($\mu=0$ and $\nu=0$ no more contributing)
\be
\sum_{\lambda=1}^3 \epsilon^{i*}(\lambda)\epsilon^j(\lambda) =\delta^{ij}
\ee
and an orthogonality relation
\be
\epsilon^i(\lambda_1) \epsilon^{i*}(\lambda_3) = - \delta_{\lambda_1\lambda_3}
\ee
with the minus sign from the spatial part of the metric.
This sign cancels the explicit sign in front of the brace of the first line of $V_d$ in  equation~(\ref{ampl1}).
Thus, the final amplitude for scattering off a heavy quark in the pion gas, to next to leading order in the chiral expansion and leading order in the heavy quark expansion, irrespective of whether the heavy quark is in a $D$ or a $D^*$ meson, is given by
\ba
V_{a} \simeq   \frac{C_{0}}{4 F^2} (s - u)  + \frac{2C_{1}m_{\pi }^2}{ F^2} h_1   +  \frac{ 2 C_{2}}{ F^2} h_3 (p_2 \cdot p_4 ) + 
\\ \nonumber 
  \frac{2 C_{3}}{ F^2} h_5 \left[ (p_1 \cdot p_2 ) (p_3 \cdot p_4 ) + (p_1 \cdot p_4 )(p_2 \cdot p_3 ) \right] \ .
 \label{amplheavy}
\ea

The $1/m_D$ pieces included in our amplitude Eq.~(\ref{ampl1}) are of course not the complete amplitude of NLO heavy-quark counting. It includes only those that are simultaneously NLO in the  chiral counting; we are dealing with a double series expansion of the total amplitude. Due to those corrections, and also for our allowing the physical $D$, $D^*$ masses to be slightly different, the properties of the $D_0$ and $D_1$ are not precisely the same. However they are close enough for most purposes.

\subsection{Unitarized scattering amplitude}

Chiral perturbation theory amplitudes are by construction a series expansion (albeit with logarithmic corrections and, in our case, Born terms with an intermediate propagator due to the $DD^*\pi$ coupling) and by their very nature are unable to describe excited elastic resonances (in our case, $D_0$ and $D_1$). \\
The key to understanding this limitation is to note that, at fixed order, ChPT violates unitarity as momentum is increased. Therefore several strategies have been adopted to bypass the shortcoming, such as the $N/D$ method, the Inverse Amplitude Method, or the K-matrix method.

We pursue the simplest partial-wave unitarization by employing on-shell factorization~\cite{Oller:1997ti} which is a nice feature of polynomial expansions and leads to algebraic formulae for the unitarized partial wave amplitudes, capable of reproducing resonances. 
Our conventions for the expansion of the perturbative $V_a$ and unitarized $T_a$ amplitudes in terms of Legendre polinomials are
\be 
V^l_a =\frac{1}{2} \int_{-1}^1 dx \  P_l (x) V_a(s,x) 
\ee
\be T^l_a =\frac{1}{2} \int_{-1}^1 dx \  P_l (x)  T_a (s,x) 
\ee
where $x \equiv \cos \theta$ and $P_0 (x)=1$ and $a$ is a channel index.

We proceed by projecting the perturbative amplitude into the $s$-wave, 
that dominates at low energies because of the $k^{2l+1}$ suppression of higher waves, and is resonant at the $D_0$ (for $D\pi$ scattering) and $D_1$ (for $D^*\pi$ scattering), thus dominating the entire amplitude at moderate heavy-quark velocities (at higher velocities, boosting to the moving center of mass frame kinematically induces higher waves).
Thus the perturbative amplitude is substituted for
\be 
V^{l=0}_a (s)= \frac{1}{2} \int_{-1}^1 dx V_a (s,t(x),u(s,t(x))) \ P_{0} (x) \ .
\ee
The unitarized scalar amplitudes $T_a$ decouple in leading order HQET
and read (Eq.~(12) of Roca, Oset and Singh~\cite{Roca:2005nm})
\be \label{Unitarizedampl}
T^{l=0}_a (s) = \frac{-V^{l=0}_a (s)}{1- V_a^{l=0} (s) \ G_{l=0} (s)}\ .
\ee
This equation manifestly is a relativistic generalization of the Lippmann-Schwinger equation.

The factorized resolvent function is the standard one-loop integral
\be 
G_{l=0} (s) = i \int \frac{d^4 q}{(2\pi)^4} \frac{1}{(P-q)^2- M_D^{2}+i\epsilon}\frac{1}{q^2 - m_{\pi}^2 + i\epsilon} \ .
\ee
We employ dimensional regularization of the divergent integral to read (Eq.~(14) from ref.~\cite{Roca:2005nm})
\begin{eqnarray}
G_{l=0} (s)= 
\\ \nonumber
\frac{1}{16 \pi^2} \left\{ a(\mu) + \ln \frac{M_D^2}{\mu^2} + \frac{m_\pi^2-M_D^2 + s}{2s} \ln \frac{m_\pi^2}{M_D^2} \right. \label{propdr} 
\\ \left.
 + \frac{q}{\sqrt{s}} \left[
\ln(s-(M_D^2-m_\pi^2)+2 q\sqrt{s})\right. \right. 
\nonumber \\ \left.\left.
+ \ln(s+(M_D^2-m_\pi^2)+2 q\sqrt{s}) \right. \right.\nonumber  
\\ \nonumber \left.\left.
+\hspace*{-0.3cm}- \ln(s-(M_D^2-m_\pi^2)-2 q\sqrt{s})
\right.\right.
\\ 
\left. \left. 
-\ln(s+(M_D^2-m_\pi^2)-2 q\sqrt{s}) -2\pi i \right] \phantom{\frac{1}{1}}
\right\} \ ,
\nonumber
\end{eqnarray}
where the  imaginary part of the logarithms above $D\pi$ threshold reads
\be 
\Im \ G_{l=0} (s) = - \frac{q}{8 \pi \sqrt{s}} \ ,
\ee
with $q$ the modulus of the pion's three-momentum in the CM frame.

Introducing the conventional two-body phase space
\be 
\rho_{ \pi D} (s)= \sqrt{\left( 1+ \frac{ (m_{\pi} + m_D )^2}{s}\right) \left(1- \frac{(m_{\pi} -m_D)^2}{s} \right)} 
\ee
or, in terms of $q$,
\be 
\rho_{ \pi D} (s)= \frac{2q}{\sqrt{s}}\ ,
\ee
this imaginary part is
\be 
\Im \ G_{l=0} (s) = - \frac{\rho_{\pi D} (s)}{16 \pi} \ .
\ee

With these ingredients it is straightforward to show that, by construction, the complex $T_a$'s satisfy single--channel unitarity relations
\be 
\Im \ T^{l=0}_a (s) = - |T^{l=0}_a(s)|^2 \frac{\rho_{\pi D} (s)}{16 \pi^2} 
\ee
(providing a convenient numerical check of our computer programmes).
The amplitude can be parametrized in terms of the phase-shift
\be 
T^{I0}_a (s)= \frac{ \sin \delta_{I0}(s) e^{i \delta_{I0}(s)}}{\rho_{\pi D}(s)} \ , 
\ee
that are then extracted via
\be 
\tan \delta_{I0} (s) = \frac{\Im \ T^{I0} (s)}{\Re \ T^{I0} (s)} 
\ee
with $I=1/2, 3/2$. 
(The tangent extraction should automatically resolve the phase-shift sign).
Finally, the isospin averaged amplitude for the LO-HQET decoupled single-channel problem becomes
\be 
|\ov{T}_a|^2 = \frac{1}{6} \left( 2 |T_a^{1/2,0}|^2 + 4 |T_a^{3/2,0}|^2 \right) \ . 
\ee
Heavy-quark spin symmetry dictates that, whether $D$ or $D^*$ in any spin state, the scattering cross-section will be the same, and since an s-wave cannot flip the spin upon interaction, no further spin averaging is needed in leading order HQET. 
One can then use
\be 
\sum | \mathcal{M}_{\pi c} (s,t,\chi)|^2 = |\ov{T}_a|^2 
\ee
in Eq.~(\ref{probdist}).

Going beyond LO in HQET we need to distinguish between $D\pi\to D\pi$ and $D^*\pi\to D^*\pi$ scattering. 
To implement it, we assume that a charm quark propagates as a linear combination of both states
\be
\ar c \ra = \alpha \ar D\ra + {\vec{\beta}}\cd \ar {\bf D}^*\ra \ .
\ee
The moduli of the complex numbers $\alpha$ and $\beta_i$ are determined by thermal Bose-Einstein distribution factors, since the mass difference between $D$ and $D^*$ slightly suppresses the latter. We then average over the relative (quasi-random) phases of $\alpha$ and ${\vec{\beta}}$ upon squaring to construct $\sum | \mathcal{M}_{\pi c} (s,t,\chi)|^2$.

For ease of comparison with other systems, we will also quote numerical results for the cross-sections given by 
\be
\sigma(s)_{\pi D} = \frac{1}{16\pi s}\ar {\mathcal{M}_{\pi D}}\ar^2
\ee
and
\be
\sigma(s)_{\pi D^*} = \frac{1}{16\pi s}\ar {\mathcal{M}_{\pi D^*}}\ar^2 \ ,
\ee
although what is substituted in the Fokker-Planck integrals is the squared matrix element of $\mathcal{M}$.

\subsection{Value of the low-energy constants}
In the philosophy of low-energy effective theories, after all the symmetries have been used to constrain the Lagrangian density, the remaining free constants have to be fit to experimental data.
Eventually these constants should also be accessible to lattice QCD.

To the order that we are working, the pion decay constant in the chiral limit $F$ can be approximated by its physical value, $f_\pi=92$ MeV, the difference being of one higher order in the chiral expansion. 

The renormalization scale for the NLO ChPT constants is to be understood as 
$\mu=770$ MeV, and the scheme is such that the subtraction constant $a(\mu)=1.85$ is fixed as in Oset, Roca and Singh~\cite{Roca:2005nm}.

The authors of Ref.~\cite{Geng:2010vw} quote a value of $g = 1177\pm 137$ MeV for the heavy-light pseudoscalar-vector coupling constant $g$, that can be obtained by reproducing the decay of $D^{ \ast +}$-mesons.
We reproduce this elementary calculation with the Lagrangian density in
Eq.~(\ref{lag1}) and obtain
\be
\Gamma = g^2 |p_{\pi}|^3/(12 \pi F^2 M^{2}_{D^*})
\ee
 in agreement with \cite{Geng:2010vw}, whose value and error band we adopt.

In his recent paper~\cite{Laine:2011is}, Laine quotes the value $g_{\pi} \sim 0.5$ for his effective Lagrangian. This Lagrangian is worked out in detail in the textbook of Manohar and Wise~\cite{Manoharstext}, where 
they quote an early value of $g_{\pi}=0.42$ from a lattice Montecarlo simulation by the UKQCD Collaboration. 

However both references employ a representation based on a heavy-hadron spinor field $H_a$ with dimension 3/2, whereas in our Lagrangian the $D$-field's dimension is 1. For these reason their $g_{\pi}$ has no dimension whereas our $g$ has dimension = 1. Direct comparison in the Lagrangian is not transparent, but instead one can easily compare the $D^*$ tree-level decay width, and find the relation among the two couplings.
The decay width employing the convention of Manohar and Wise reads 
$$
\Gamma = g^2_{\pi} |p_{\pi}|^3/(6 \pi F^2)
$$
yielding $g=\sqrt{2} g_{\pi} M_{D^*} \sim 1190$ MeV, in agreement with the value from~\cite{Geng:2010vw}.

Turning now to the $NLO$ constants, we have repeatedly stated that $\tilde{h}_i=h_i$ is a requirement of heavy quark symmetry tying the $D$ and $D^*$ amplitudes at LO in Heavy Quark Effective Theory. We, in this article, set therefore $\tilde{h}_i-h_i=0$ from the start. Likewise we have set $h_0=h_2=h_4=0$ based purely on large-$N_c$ counting. These constants well deserve being revisited in future work, but we are content here with accepting a $1/N_c$ systematic error as customary in the current literature.

Another useful constraint is offered by the mass differences between the $D$-mesons \cite{Lutz:2007sk,Geng:2010vw}, which fixes $h_1 \approx -0.45$. 
Thus, the remaining free LECs to be estimated are $h_3$ and $h_5$. 
We have at our disposal, in the $D\pi$ channel corresponding to the $T_a$ scattering amplitude, two pieces of known data (the $D_0$ mass and width) to which we can tie $h_3$ and $h_5$. If a calculation including the subleading order in HQET is performed, then the $D_1$ and $D_0$ parameters differ and the constants become overconstrained by known data.
We will find in section~\ref{sec:numerics} that reasonable values are
$(h_3,h_5)=(7\pm 2,-0.5\pm 0.2 \textrm{ GeV}^{-2})$ with correlated errors, that is, an increased $h_3$ needs to be used with a more negative $h_5$.

A word of caution seems convenient about the numerical value of $h_3$ and $h_5$.
In~\cite{Guo:2009ct} it has been proposed that $h_5 = (h'_5/m_D^2) \sim \mathcal{O}(1/m^2_D)$ since $h'_5$ is assumed there to be of order 1. However we think this is unnaturally small and that $h_5=\mathcal{O}(g^2/\Lambda^2_{\rm QCD})$ should be expected.
Our reasoning is based on resonance saturation. Instead of unitarizing the amplitude and fitting the constants to the dynamically generated resonances, we could have introduced the resonances as additional fields and eliminated them from the low-energy theory~\cite{Ecker:1988te}
by employing 
$$
g \frac{-i}{p^2-m_{D_0}^2}g \to \frac{ig^2}{(m_D+m_\pi)^2-m_{D_0}^2} 
$$
near threshold.
It is clear that the denominator is proportional to the off-shellness of the $D_0$ resonance and not to its total mass. Of course, the analogous quantities coincide in the traditional case of $\pi\pi$ scattering since pions are so light as compared to the $\rho$ for example, $m_\pi=138$ MeV $\ll m_\rho=770$ MeV. Then $p_{\pi\pi}\simeq 0$. However the mass of the ground-state charmed meson cannot be neglected in $D$-pion scattering and the low-energy constants do not vanish in the heavy-quark limit.
Thus we would expect the denominator to be of order $\Lambda_{\rm QCD}^2$ or at most
$m_\pi m_D$, but not $m_D^2$. 
Some additional discussion about the $h_i$ constants can be found in a recent paper in the heavy quark limit~\cite{Cleven:2010aw}.

\section{Numerical results} \label{sec:numerics}
\subsection{Cross section for $D\pi$ elastic scattering}

We now present numeric computations of the unitarized and squared amplitudes in Eq.~(\ref{Unitarizedampl}), and of the cross-section.

\begin{figure}
\centering
\includegraphics[width=8cm,trim = 0mm 0mm 0mm 1mm, clip]{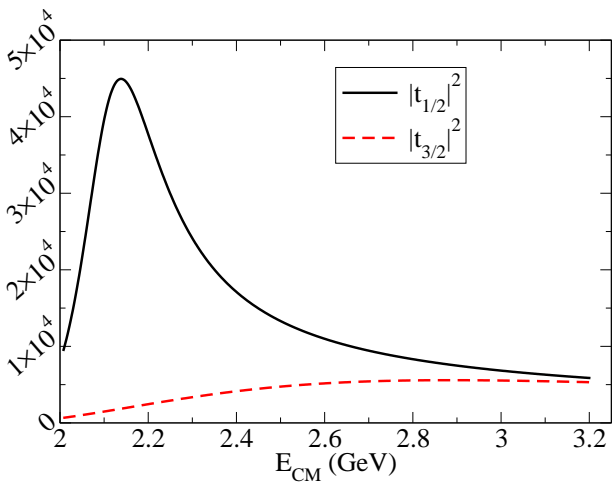}\\
\vspace{1cm}
\includegraphics[width=8cm,trim = 0mm 0mm 0mm 1mm, clip]{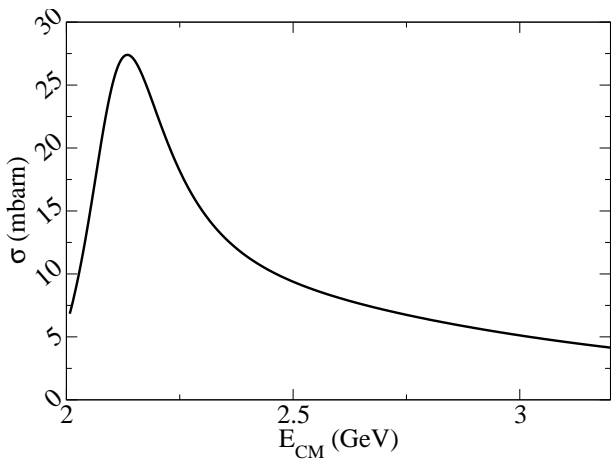}
\caption{\label{fig:sminusuonly} 
Top: square amplitudes for $D\pi$ $s$-wave elastic scattering employing only the $(s-u)$ term of the interaction potential $V_a$ (as in Gamermann and Oset).
Bottom: isospin averaged cross section associated to those amplitudes.
}
\end{figure}

In the first place, and to compare with the work of Gamermann and Oset~\cite{Gamermann:2007fi}, we keep only the $(s-u)$ term in the $D\pi$ elastic amplitude $V_a$. The square amplitudes with isospin $I=1/2$ and $I=3/2$
and $l=0$ are depicted in Fig.~\ref{fig:sminusuonly}.
The figure shows how the exotic $I=3/2$ is non-resonant (this will also be the case for all the calculations presented below), which could have been guessed because no $q\ov{q}$ state exists with such isospin, so there is no intrinsic strength at low energies in exotic waves.
The non-exotic $I=1/2$ channel presents a clear $s$-wave resonance, with approximate mass and width $M\simeq 2140$ MeV and $\Gamma\simeq 170$ MeV. These values are somewhat too low if compared with the experimental $M_{D_0}=2360(40)$ MeV and $\Gamma_{D_0}=270(50)$ MeV taken from the Review of Particle Physics. 

We do not deem this a problem since there is room for the $NLO$ terms containing the $h_i$ constants to modify the computation.
But how can then Gamermann and Oset obtain reasonable agreement with the experimental state, employing only the leading order amplitude in ChPT?

We believe to have identified the reason in their substituting one of the powers of $f_\pi$ by  $f_D$, 
$$
\frac{C_0 (s-u)}{4f_\pi^2} \to \frac{C_0 (s-u)}{4f_\pi f_D} \ .
$$
This suppresses the strength of $V_a$ such that $T_a$ saturates unitarity at a higher center of mass-energy $\sqrt{s}$, in better agreement with experimental data. Reducing $V_a$ by a factor 2 displaces the maximum of the cross section to about $2320$ MeV with width about $250$ MeV.

The substitution of $f_\pi$ by $f_D$ can be tracked to those authors employing $SU(4)$ symmetry to construct the effective Lagrangian, treating $D$ mesons on equal footing with pions. However we believe this is a questionable procedure since $SU(4)$ is not even an approximate symmetry, and we have instead constructed the chiral coupling of pions to the heavy $D$-meson source. 

The $LO$ interaction is therefore somewhat too strong in our case. We could weaken it by modifying the subtraction constant of the loop function $a(\mu)$, but instead we proceed to the next order in chiral perturbation theory, which should be equivalent as shifts in $a(\mu)$ should be absorbed in the $NLO$ $h_i$ constants.

Next we add one by one the NLO constants $h_1$, $h_3$, $h_5$.
Because the $h_1$ term does not increase with momentum, but is multiplied by a small $m_\pi^2$ constant, it does not change the amplitudes appreciably. We include it but do not discuss it any further.

\begin{figure}
\centering
\includegraphics[width=8.5cm,trim = 0mm 0mm 0mm 1mm, clip]{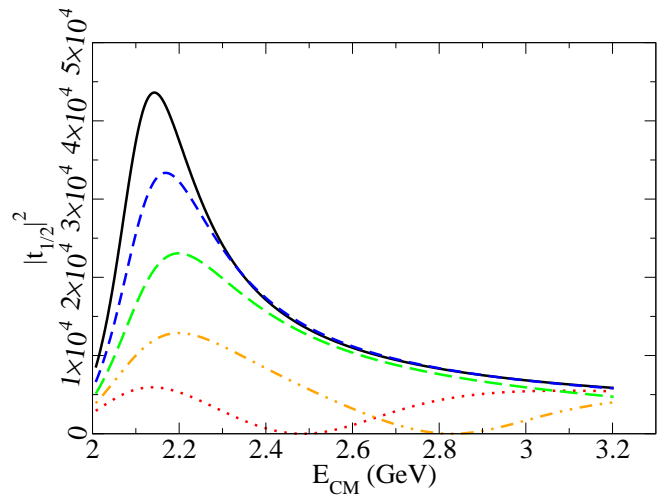}
\caption{\label{fig:withh3} 
Squared isospin $1/2$ amplitude for $D\pi$ scattering for various values of $h_3$, from top to bottom being 0,1,2,3,4. In this graph $h_5$ is kept fixed at zero.\vspace{0.3cm}
}
\end{figure}

We examine then the sensitivity to $h_3$ in Fig.~\ref{fig:withh3}. 
For small, positive values of $h_3$ the $D_0$ peak moves to larger masses, with $h_3=2$ the shift is of order 50 MeV. The resonance also becomes broader. Then, for larger values $h_3\ge 3$, the mass starts falling again, and a cancellation with the $s-u$ term sets in, forcing a zero of the amplitude at energies 2.5 GeV or above.

If we now add the $h_5$ term, we observe that its presence (if the sign is chosen negative as in Guo et al., for example $h_5=-0.25$ GeV$^{-2}$) narrows the resonance shifting it to slightly lower masses. If positive, $h_5$ forces a cancellation (as did a large $h_3$) giving a zero near threshold (for $h_5\simeq 1$ GeV$^{-2}$, $h_3\simeq 2$) or at $2.4$ GeV and above (for the same $h_3$ but $h_5\simeq 0.25-0.5$ GeV$^{-2}$).

Therefore a strategy to improve agreement with the experimental $D_0$ data is to combine a positive $h_3$ with a negative $h_5$ to increase the resonance mass without distorting the line-shape unacceptably.

\begin{figure}
\centering
\includegraphics[width=8.5cm]{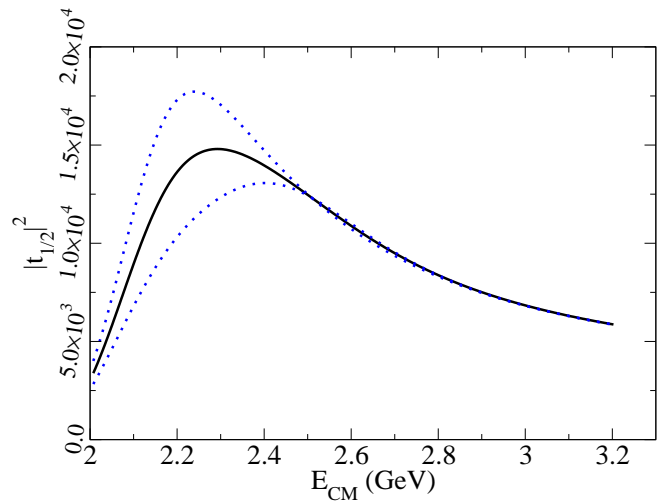}
\caption{\label{fig:withh5} 
Squared isospin $1/2$ amplitude for $D\pi$ scattering with $(h_3,h_5)=(7,-0.5$ GeV$^{-2})$ (central value). 
}
\end{figure}

Our best computation is then shown in Fig.~\ref{fig:withh5}.
 
The maximum of the squared amplitude, employing $(h_3,h_5)=(7,-0.5$ GeV$^{-2})$ as central value, gives a reasonable $M_{D_0}=2300$ MeV, just slightly below the experimental value, and a width just slightly above $\Gamma=350$ MeV.
The two parameters are very correlated, so that varying one significantly requires varying the other simultaneously to maintain reasonable agreement with the experimental resonance. Shown in the figure are two more lines with the error band $\Delta h_3=\pm 2$ and $\Delta h_5 = \pm 0.2$ GeV$^{-2}$.
It is this squared amplitude, leading order in Heavy Quark Effective Theory,  that we adopt in our Fokker-Planck equation for the transport coefficients.

\begin{figure}
\centering
\includegraphics[width=8.5cm,trim = 0mm 0mm 0mm 1mm, clip]{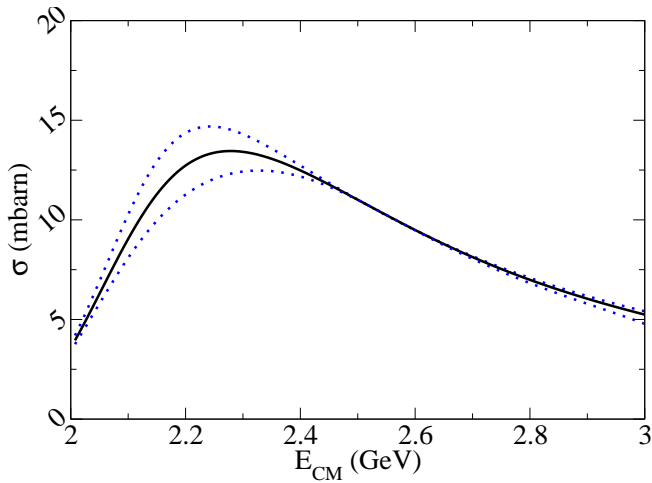}
\caption{\label{fig:withh5cross} 
Cross section for $D\pi$ elastic scattering with $(h_3,h_5)=(7,-0.5$ GeV$^{-2})$ (central value). 
}
\end{figure}

Although the diffusion and drag coefficients require the $\ar \mathcal M \ar^2$ square amplitude, it is convenient for the discussion to also plot the resulting cross section, which we do in Fig.~\ref{fig:withh5cross}.

The maximum of the cross-section is about $13.5\pm 1$ mbarn, and for the entire range of center of mass energies $\sqrt{s}\in(2-3)$ GeV we find 
$\sigma\ge 5$ mbarn. In fact, for a large window between 2.1 and 2.5 GeV we have $\sigma \ge 10$ mbarn, which is slightly larger but in reasonable agreement with the guess by the authors of~\cite{He:2011yi}, that assume $7-10$ mbarn, or by Svetitsky and Uziel~\cite{Svetitsky:1996nj} of 9 mbarn.

\begin{figure}
\centering
\includegraphics[width=8.cm,trim = 0mm 0mm 0mm 1mm, clip]{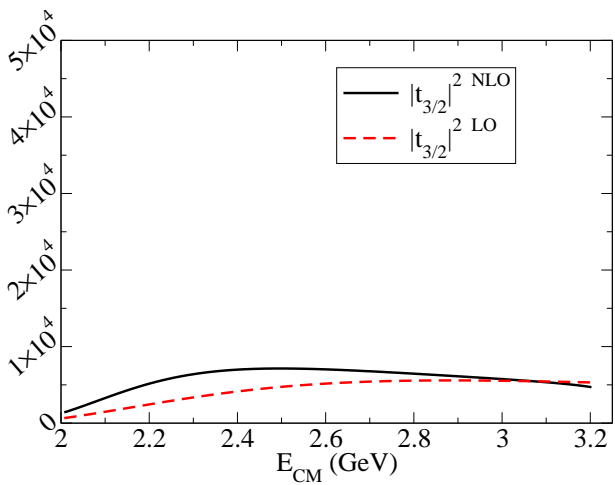}
\caption{\label{fig:is3halves} 
Effect of adding the $h_i$ counterterms to the $s-u$ basic $D\pi$ amplitude for isospin $3/2$. \vspace{0.3cm}
}
\end{figure}

For the sake of completeness, we separately quote the effect of adding the $h_i$ constants on the non-resonant isospin $3/2$ $D\pi$ elastic amplitude.
The corresponding plot is number~\ref{fig:is3halves}.

As can be seen in the figure, the effect is moderate at all energies.

\begin{figure}
\centering
\includegraphics[width=8.cm,trim = 0mm 0mm 0mm 1mm, clip]{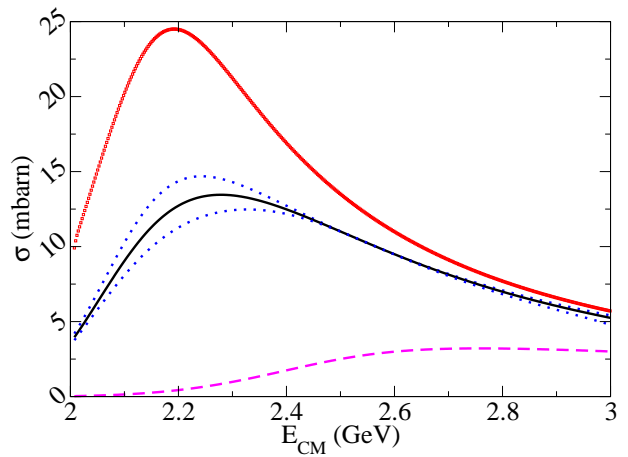}
\caption{\label{fig:Born1} 
Effect of including the Born terms associated with the $D^*$. 
The bottom line (purple) is the cross-section associated to the Born term alone, as in the model of~\cite{Ghosh:2011bw}. The top line (red squares) is the resulting cross-section combining the Born term with the contact terms, without modifying the $h_i$ constants from Fig.~\ref{fig:withh5cross}, and then unitarizing. 
}
\end{figure}

Next, we proceed to the next-to-leading order in Heavy Quark Effective Theory.
We only consider for now the Born $s$ and $t$-channel exchange terms due to $D^*$ exchange between the $D\pi$ pair. \\
The effect of adding these terms is akin to making $h_5$ more negative, that is, a narrowing of the $D_0$ resonance, as shown in Fig.~\ref{fig:Born1}. 

\begin{figure}
\centering
\includegraphics[width=8.cm,trim = 0mm 0mm 0mm 1mm, clip]{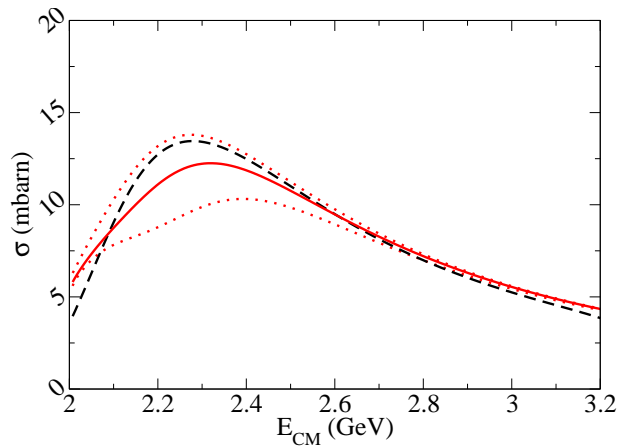}
\caption{\label{fig:Born2} 
Effect of including the Born terms associated with the $D^*$, but leaving the $h_i$ coefficients free. The red, solid line is the central value with $h_3=8$, $h_5=0.35$ GeV$^{-2}$. The black, dashed line coincides with the cross section in Fig.~\ref{fig:withh5cross} without the Born terms.
}
\end{figure}

However, a renormalization of the $h_i$ constants effectively brings back the pole position in better agreement with experimental data. We now refer the reader to Fig.~\ref{fig:Born2}. 
Shown in the figure are lines with $(h_3,h_5)=(7.5\pm 2.5, 0.4\pm 0.3$ GeV$^{-2})$, together with the result of Fig.~\ref{fig:withh5cross} without including the Born terms. As can be seen, the effect of the $D^*$ exchanges can be largely absorbed in the $h_i$ counterterms (for fixed $m_c$ mass of course, since they scale differently) so we will ignore the Born terms in this computation. However a certain uncertainty should be understood, of order 30\% in the cross section , that could be larger than our estimate in the region of the $D_0$ resonance.

It is also worth commenting that the addition of the Born terms causes a dip in the high-mass $D\pi$ spectrum that can be brought down by minimum changes in the $h_i$ parameters. Since we do not think that such an interference dip between $D^*$ Born exchange and the $D_0$ tail has been reported in experimental data, we keep the contact parameters in a band such as not seeing this dip in the momentum range of relevance.

\begin{figure}
\centering
\includegraphics[width=8.cm,trim = 0mm 0mm 0mm 1mm, clip]{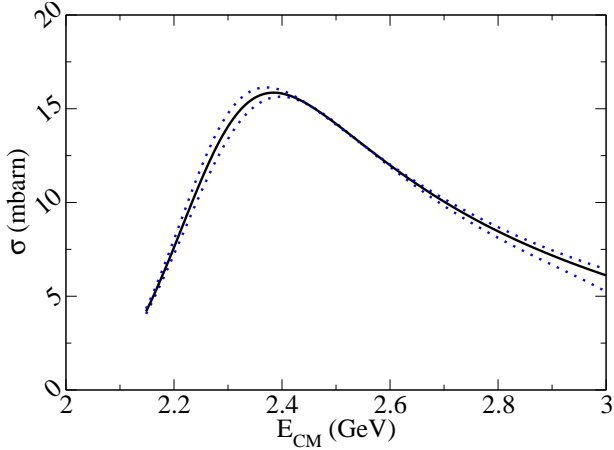}
\caption{\label{fig:Dstarscattering} 
Elastic cross-section for $D^*\pi$ scattering computed replacing $m_D$ by
$m_D^*$ in Fig.~\ref{fig:withh5cross}. The resonance should now be interpreted as the broad $D_1(2427)$.
}
\end{figure}

Finally, we return to the computation in Fig.~\ref{fig:withh5cross}, but substitute $m_D$ by $m_D^*$ (an NLO effect in HQET) as only modification to obtain $V_d$ instead of $V_a$. We interpret the resulting cross-section as 
that corresponding to $D^*\pi$ scattering, and plot the result in Fig.~\ref{fig:Dstarscattering}.
The cross-section including both $1/2$ and $3/2$ isospin channels is clearly resonant, with the $D_1$ well visible. As was the case for the $D_0$,
the mass is slightly below the data.
The cross-section peak is about 15 mbarn.

Thus we have performed an exhaustive study of the LO-HQET interaction and now proceed to compute transport coefficients equipped with the interaction leading to Figs.~\ref{fig:withh5cross} and~\ref{fig:Dstarscattering}.

\subsection{Diffusion and drag coefficients}\label{subsec:numtransport}

\begin{figure}
\centering
\includegraphics[width=8.5cm]{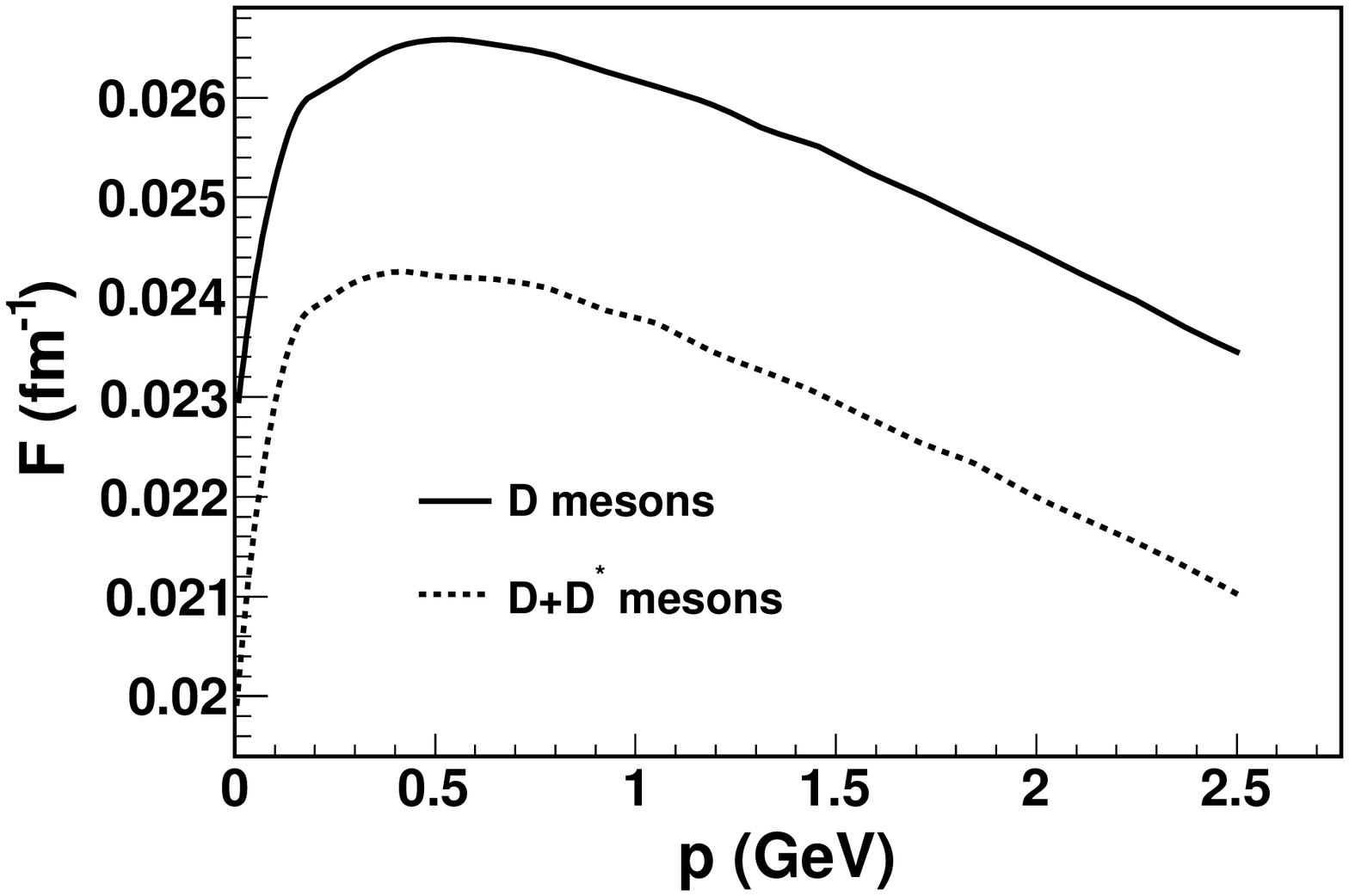}\\
\includegraphics[width=8.5cm]{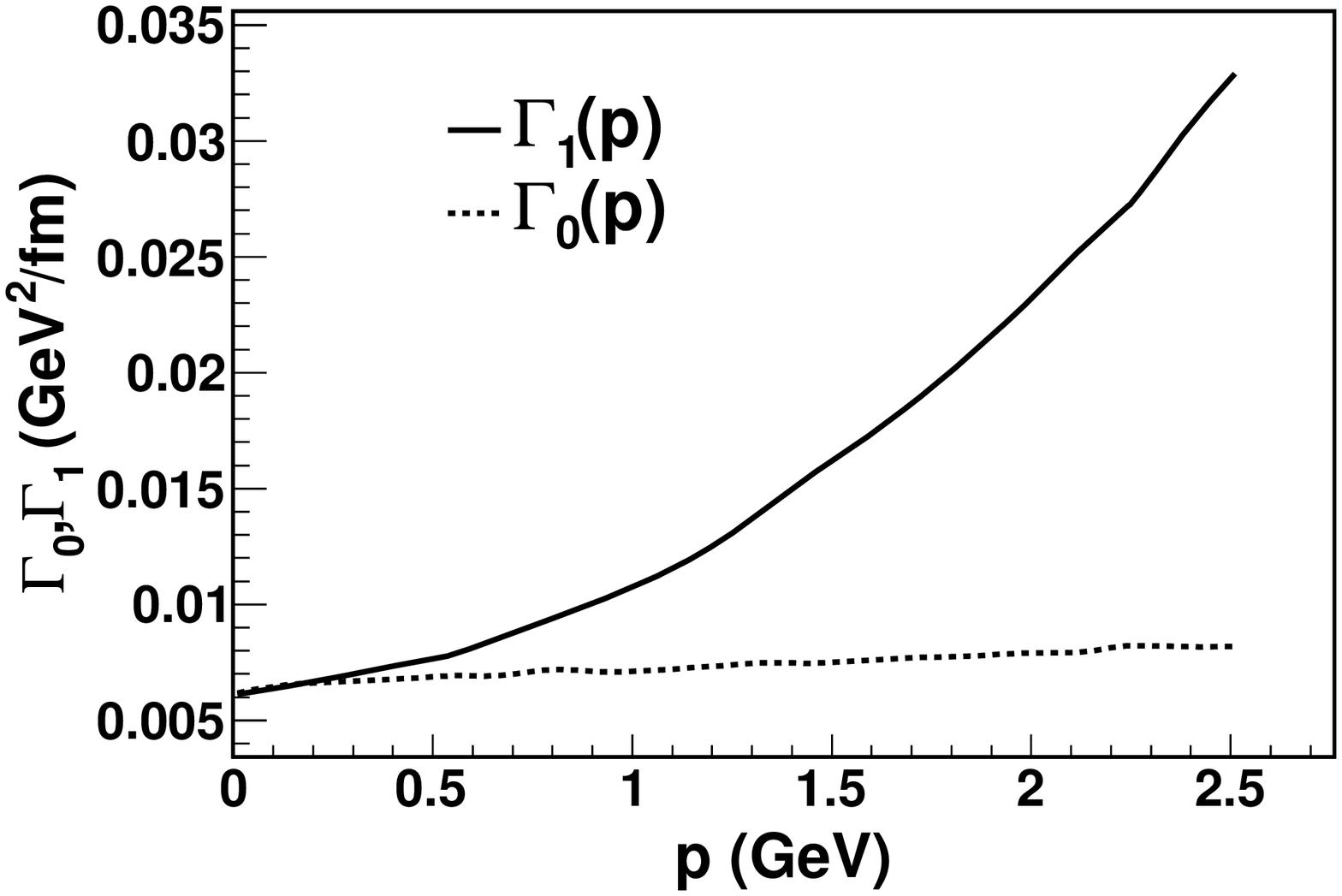}\\
\includegraphics[width=8.5cm]{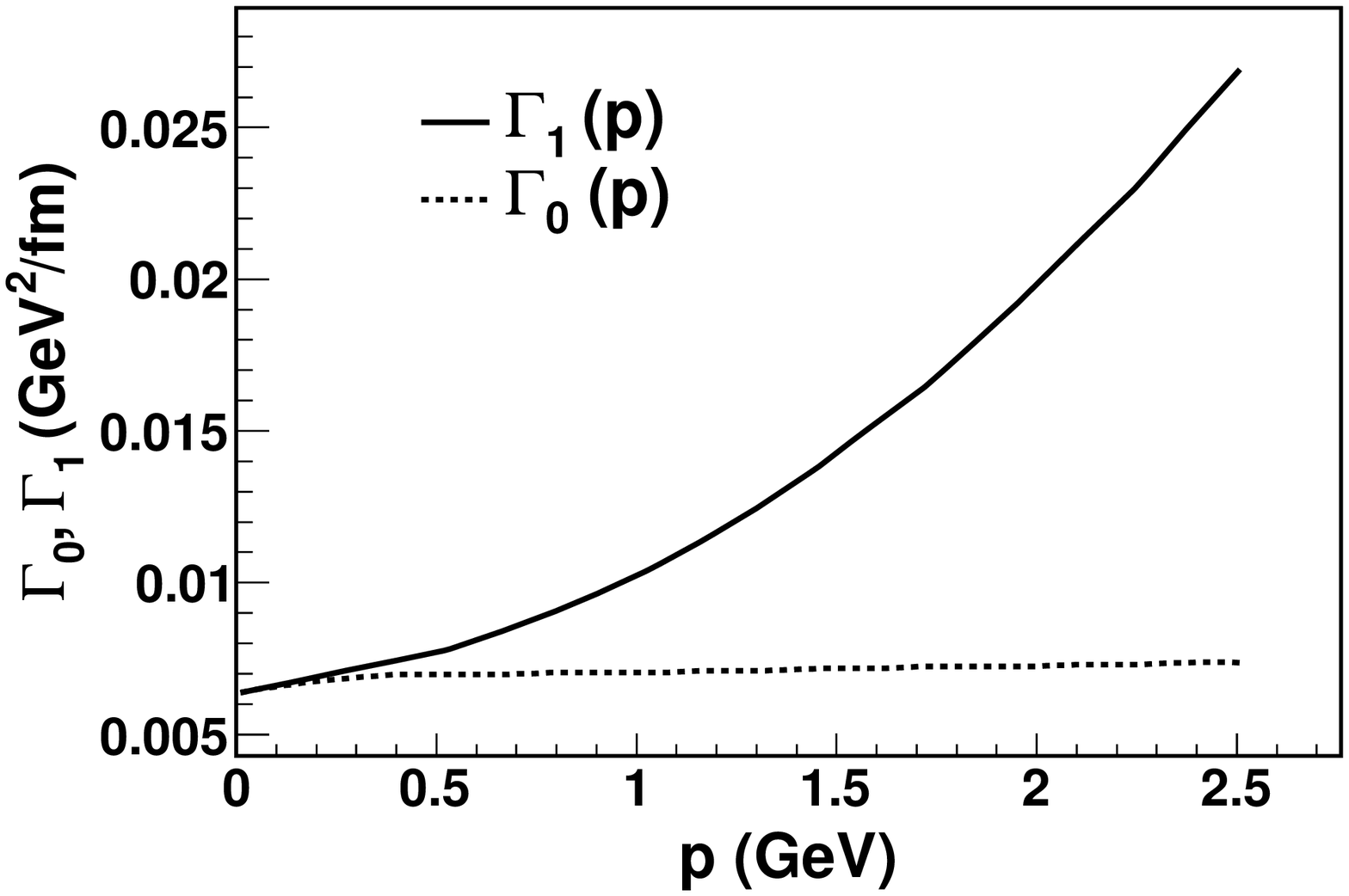}
\caption{\label{fig:threcoeffs} 
We show all three coefficients in the Fokker-Planck equation as a function of charm-quark momentum, at a reference temperature of 150 MeV in the pion gas.
The low-energy constants in the $D \pi \rightarrow D \pi$ amplitude are fixed to $h_1 = - 0.45$, $g=1177$ MeV, and $h_3$ and $h_5$ fit to describe the mass and width of the $D_0$ resonance. Top: $F$ including and not including the possible propagation of the $c$ quark as a $D^*$ meson. Middle: $\Gamma_0$ and $\Gamma_1$
including $D$-like propagation alone. Bottom: $\Gamma_0$ and $\Gamma_1$ including also propagation as a $D^*$ meson.
}
\end{figure}

We now proceed to computing, with the square amplitude so numerically computed, the $F$, $\Gamma_0$ and $\Gamma_1$ transport coefficients.
The three pannels of Fig.~\ref{fig:threcoeffs} shows them as function of squared momentum $p^2$ for fixed temperature $T=150$ MeV. One should not trust these results above charm momenta of order $p=1.5$ GeV, but we spell them out for completeness.

In the top panel of this figure we show the drag coefficient $F(p)$ in fm$^{-1}$, which exhibits a momentum dependence of about 10 \% within the range of $p\in(0,2.5)$ GeV.
From this coefficient one can extract the relaxation length for a charm quark propagating in the pion medium that turns out to be around $40$ fm at $p=$1 GeV.

Quite strikingly, one can see in the figure that $\Gamma_0$ has a very mild momentum dependence, its value can very well be approximated by a constant for the entire momentum range. $\Gamma_1$ is seen to grow with momentum, increasing the difference $\Gamma_1-\Gamma_0$, and thus favoring diffusion at higher typical momenta.

\begin{figure}
\centering
\includegraphics[width=8.5cm]{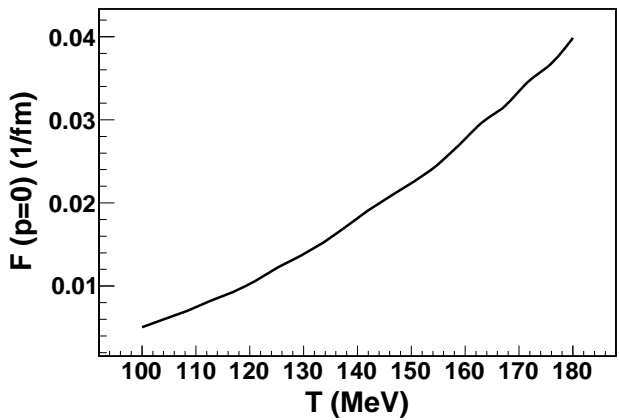}
\caption{\label{fig:dragfixedP0} 
Momentum-space drag coefficient as function of temperature for a stopped charm quark in the hadron gas. We obtain the coefficient by employing the Einstein relation in taking the limit of $p\to 0$.}
\end{figure}

\begin{figure}
\centering
\includegraphics[width=8.5cm]{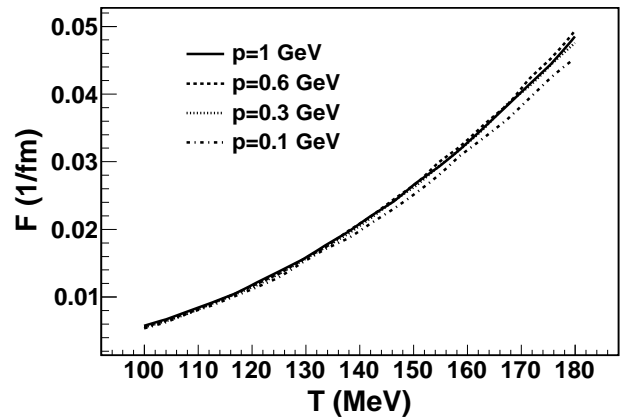}
\includegraphics[width=8.5cm]{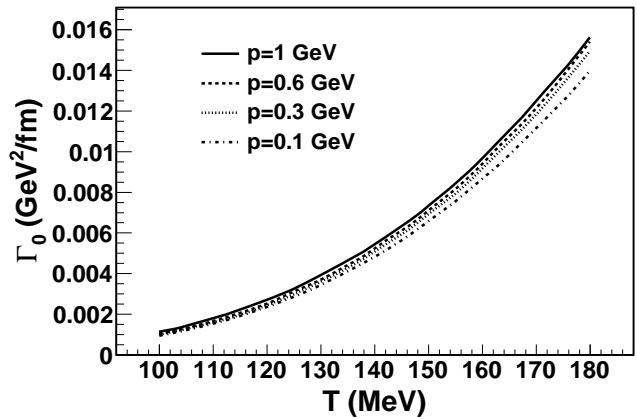}
\includegraphics[width=8.5cm]{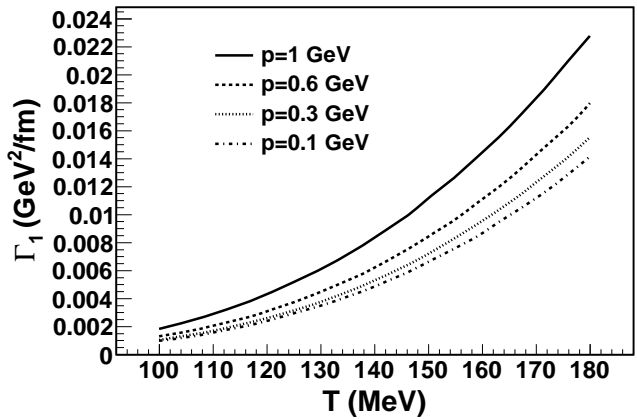}
\caption{\label{fig:dragfixedP} 
Momentum-space drag and diffusion coefficients as function of temperature for a slow charm quark with momentum $p=1$ GeV, 0.6 GeV, 0.3 GeV and 0.1 GeV.
Note that the intensity of the drag force is roughly proportional to the temperature. 
}
\end{figure}

In Figs. \ref{fig:dragfixedP0} and \ref{fig:dragfixedP} we show the dependence with the temperature of the drag coefficient at fixed momentum. Since the direct computation of $F(p^2\to 0)$ is rather unstable, the plot in Fig.~\ref{fig:dragfixedP0} is computed from $\Gamma$ by employing the Einstein relation, Eq.~(\ref{Einstein}) in the Appendix. 

The drag coefficient is seen to increase by a factor of about 4 in the range from 100 to 150 MeV, so that most of the drag in a heavy ion collision is expected in the hotter stages, with the charm quarks freezing out progressively until they freely stream outwards till they decay.

We compare with other authors, choosing a reference temperature of 100 MeV where all existing works make a statement, and show the drag coefficient for each recent work in Table~\ref{tab:drag}.
\begin{table}[h]
\caption{Value of the drag coefficient at $p\rightarrow 0$ and $T=100$ MeV. \label{tab:drag}}
\begin{center}
\begin{tabular}{|cc|} \hline
Authors             & $F(\textrm{fm}^{-1}$) \\ \hline \hline
Laine               & $0.05\times 10^{-3}$ \\ 
He, Fries, Rapp     & $5\times 10^{-3}$ \\
Ghosh {\it{et al.}} & 0.11 \\
This work           & ${\mathbf{3.5\times 10^{-3}}}$ \\
\hline
\end{tabular}
\end{center}
\end{table}
It can be seen that the phenomenological model of He, Fries and Rapp is of the same order of magnitude of our result, with Ghosh {\it{et al.}} quoting an extremely large value in their Fig.~2, and Laine a smaller value by one order of magnitude. We believe that we have a larger control of the charm-pion scattering amplitudes at moderate temperatures, but the reader would be cautious to employ a factor 2 as error band to our result.

\begin{figure}
\centering
\includegraphics[width=8.5cm]{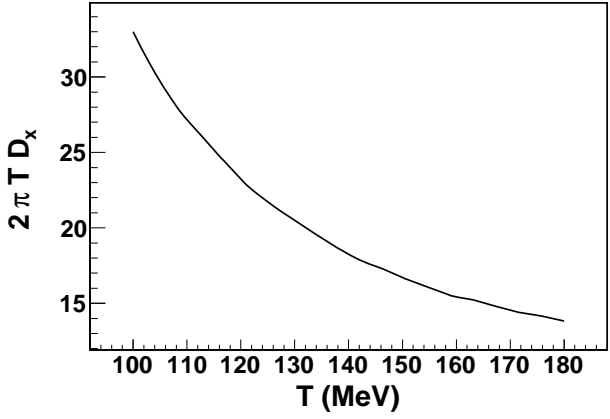}
\caption{\label{fig:spacediff} 
Spatial diffusion coefficient as a function of temperature.
}
\end{figure}

The spatial diffusion coefficient is then plotted in Fig.~\ref{fig:spacediff}
as function of temperature.

\begin{figure}
\centering
\includegraphics[width=8.5cm]{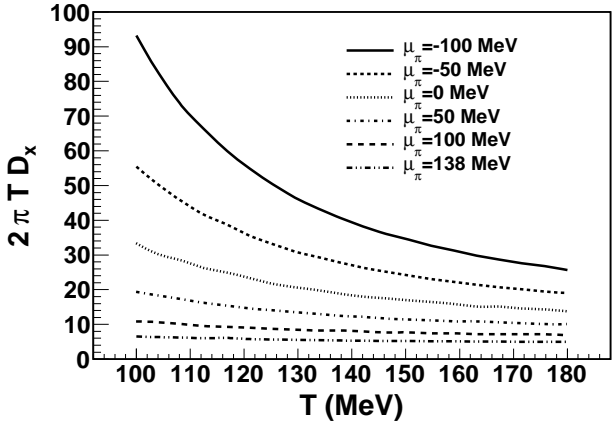}
\caption{\label{fig:spacediff_mu} 
Same as in Fig.~\ref{fig:spacediff} but as a function of the chemical potential 
}
\end{figure}

At low temperatures it correctly takes the non-relativistic limit 
\be
D_x = \frac{3T^{3/2}}{\sigma P\sqrt{m}}
\ee
with $m$ the particle mass, $\sigma$ the cross section, and $P$ the pion gas pressure, that is temperature dependent.
We also note that, during the lifetime of the pion gas after the cross-over from the quark-gluon plasma phase, the interactions between pions are almost entirely elastic, so that pion number is effectively conserved and one should introduce a pion chemical potential, not included in the very recent works by other groups. 
Introducing this approximate pion chemical potential $\mu$,
\be
P\propto m_\pi^{3/2} T^{5/2} e^{\frac{\mu-m_\pi}{T}}
\ee 
makes the product $TD_x$ diverge at low temperature and vanishing chemical potential
(which just means that gas particles are too cold and slow to stop the charm quark from diffusively moving inside the pion gas).
However, at chemical equilibrium with $\mu\to m_\pi$ (that is not expected in the hadron phase of a heavy-ion collision, but is relevant to make contact with the non-relativistic limit), the exponential becomes unity and $TD_x$ becomes a constant at low temperature.
We further show the effect of this pion chemical potential in Fig.~\ref{fig:spacediff_mu}.

We find the effect sizeable. At a reference temperature of 120 MeV, the ratio between $D_x$ at $\mu_\pi=0$ and $\mu_\pi=138$ MeV is a factor of about 5.

To assist in the physical interpretation of these results, we have plotted in Figs.
\ref{fig:energyloss} and \ref{fig:momentumloss} the loss of energy and momentum per unit length discussed in Appendix~\ref{appendixFDR} and derived from our results for the drag coefficient $F$, for various momenta $p^2$. 
\begin{figure}
\centering
\includegraphics[width=8.5cm]{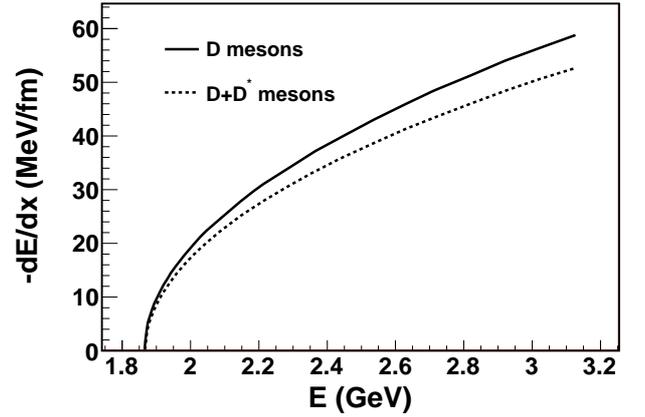}
\caption{\label{fig:energyloss} 
Loss of energy of a charmed meson as function of the energy in a pion gas at a fixed temperature of 150 MeV, assuming it can travel as a $D$ or a $D^*$ meson during the few fermi of the gas's lifetime.
}
\end{figure}
\begin{figure}
\centering
\includegraphics[width=8.5cm]{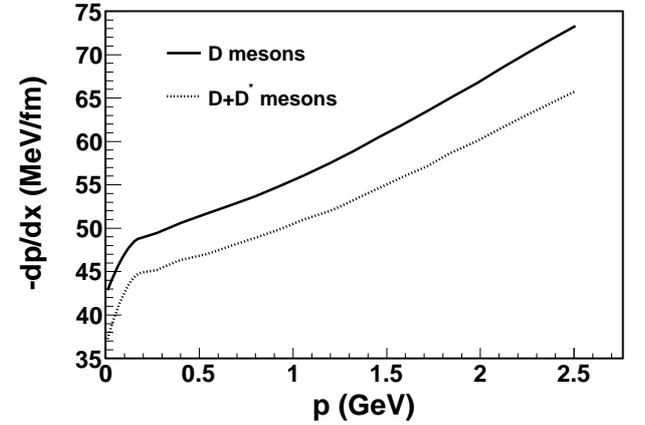}
\caption{\label{fig:momentumloss} 
Loss of momentum per unit length as function of momentum of a charmed meson in a pion gas, same as in Fig.~\ref{fig:energyloss}.
}
\end{figure}

From Fig.~\ref{fig:momentumloss} one can estimate that a reference charm quark in a $D$ or $D^*$ meson with momentum 1 GeV measured in the rest frame of the pion fluid surrounding it, will deposit about 50 MeV per Fermi travelled in the fluid. Thus, if the pion gas is in existence for, say, 4 fm, the $D$ meson measured in the final state with a momentum
of 800 MeV will have been emitted from the quark-gluon plasma phase with a GeV. This result is similar to the $20\%$ effect recently quoted by He, Fries and Rapp~\cite{He:2011yi} and means that, while the $D$ and $D^*$ mesons can be used as probes of the quark-gluon plasma, their distributions should be shifted up in momentum (or alternatively both the quark-gluon plasma and hadron phases have to be treated in hydrodynamic simulations).

\begin{figure}
\centering
\includegraphics[width=8.5cm]{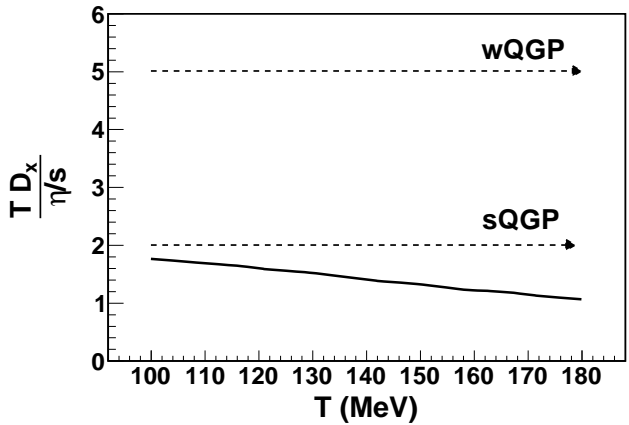}
\caption{\label{fig:stronglycoupled} 
A dimensionless ratio with the viscosity over entropy density, proposed in Ref.~\cite{vanHees:2005wb}. The top dashed line corresponds to a ``weakly coupled'' quark-gluon plasma, the bottom line to a ``strongly coupled'' quark-gluon plasma. The solid line at the bottom, for charm propagating in our pion gas, is more suggestive of the second than of the first.
}
\end{figure}

The authors of reference~\cite{vanHees:2005wb} proposed to divide the temperature times the spatial diffusion coefficient by the shear viscosity over entropy density ratio $\eta/s$, producing a dimensionless quantity that should give an idea of how strongly coupled is the quark-gluon plasma, and they quote two estimates based on AdS-CFT that we plot in Fig.~\ref{fig:stronglycoupled}.
In the figure we also plot our computation based on charm quarks travelling through the pion gas, together with our computation of viscosity over entropy density in the pion gas presented in~\cite{Dobado:2008vt}. It seems that, according to this criterion, the charm quark is somewhat strongly coupled to the pion gas, although it is not clear what the precise value of these AdS-CFT based estimates is.

\section{Experimental discussion}\label{sec:exp}

While we are not directly computing the experimentally observed quantities in this article, it is worth looking ahead onto what impact our results have for the Heavy-Ion collision programme at RHIC and at the LHC.

A commonly quoted observable is the nuclear suppression factor $R_{AA}$ obtained by dividing the number of electrons from heavy meson decays in
a nucleus-nucleus collision by the number in proton-proton collisions times the number of constituent nucleons,
$$
R_{AA} = \frac{N_{AA}}{A^2\times N_{pp}}\ .
$$
At high $p_T$ of order 5-7 GeV this ratio reaches 0.3, showing substantial effects due to the medium.
At low transverse momentum up to to 2 GeV this suppression factor 
is close to 1 (small effect). 
At the lowest $p_T$ of few hundred MeV the ratio is even larger than unity, there being an enhancement of the number of heavy mesons in ion-ion collisions~\cite{Averbeck:2008zz} over proton-proton. This pile-up of heavy mesons at low momentum can be interpreted as they being slowed down by the medium. 
The $F$ and $\Gamma_1$ coefficients are relevant for this process, with $\Gamma_1$ broadening the $p_T$ distribution and $F$ equating the velocity of the heavy quarks to the velocity of the fluid medium that they are crossing, as can be seen in Eq.~(\ref{Rayleighs}) below.

Another important observable is the elliptic flow~\cite{Gombeaud:2007ub}
 $\nu_2$ defined by the distribution of particles with the azimuthal angle $\phi$ taken around the collision axis, with the collision plane at $\phi=0$,
$$
\frac{1}{N}\frac{dN}{d\phi} = \frac{1}{2\pi} \left(
1+2\nu_1\cos \phi +2\nu_2\cos 2\phi + \dots \right) \ .
$$
Substantial elliptic flow for heavy flavored mesons has been measured at RHIC~\cite{Averbeck:2008zz}, meaning that the heavy quarks are partly equilibrating with the medium. This elliptic flow can potentially provide sensitivity to the combination of diffusion coefficients $\Gamma_1-\Gamma_0$.
In subsection~\ref{subsec:numtransport} we showed that this difference grows with quark momentum, which may help explain why the elliptic flow grows in the $p_T$ 0-2 GeV range.

As for the actual spectrum of $D$-mesons deduced by STAR~\cite{LaPointe:2010zz}, although the errors are very large, they quote an average velocity of $\beta=0.35-0.47$, sufficiently smaller than 1 to make HQET a reasonable starting point, especially taking into account that part of this velocity is due to the local fluid rest frame being in motion in the laboratory frame, with the actual charm velocity respect to that Eulerian frame being even smaller.
The STAR collaboration also quotes a rough temperature of 120 MeV as fitting their spectrum, but given that they have only three points in the plot and the large error bars, a larger temperature (or a poor thermal fit) would not be surprising at all. We should wait for future data to clarify this point.

These observables have not yet been provided by direct reconstruction of the $D$ or $B$ mesons (maybe ALICE can provide a measurement with its Internal Tracking System assisting the secondary vertex reconstruction), but indirectly with measurements of the (presumed) secondary electrons from heavy flavor decays. Copious $D$-meson counts have already been informally reported in the $K\pi$ and $K\pi\pi$ channel and we look forward to the publication of this data.

 However we feel that the measurements are very encouraging and that we should expect these charm drag and diffusion coefficients to become accessible. Then it will be necessary to disentangle the conventional diffusion in the hadron phase from the more exotic quark and gluon phases, and our results will be useful here insofar as they greatly reduce the uncertainty in the hadron, low temperature phase.

In Fig.~\ref{fig_1dsol} of the appendix we show the solution to a one-dimensional version of the Fokker-Planck equation that comes handy  for this discussion (assuming mid-rapidity and no azimuthal flow, the charm diffusion, although more complicated, is reminiscent of that one-dimensional case). 

From that figure one can see that if the initial distribution of charm quarks would peak at some $p_0\in(1,2)$ GeV, for every femtometer spent in the pion gas, the charm quark distribution would peak 50 MeV lower (friction), and the distribution would be about 100 MeV broader (diffusion).

In fact, the ALICE collaboration has already published an analysis for proton-proton collisions to serve as benchmark~\cite{dainese} for what is to come in Pb-Pb.
As usual, the mid-rapidity $p_t$ $D$-meson spectrum has an exponential shape
\be
\left[ \frac{dN}{dp_t}\right]_{y<0.5} \propto e^{-\frac{p_t}{\Lambda}}
\ee
with a scale $\Lambda\simeq 1.5$ GeV and it will be interesting to convolute our Fokker-Planck kernel with this input experimental spectrum.

\section{Summary and conclusions}

If we take 150 MeV as the highest temperature at which our approach is reliable (as we do not include strange mesons), and we examine a charm quark travelling as a meson with momentum 1 GeV, the relaxation length read off from the top plot in Fig.~\ref{fig:threcoeffs} is
$$
\lambda_{c}(p=1\textrm{ GeV},T=150\textrm{ MeV}) \simeq \frac{1}{0.025\textrm{ fm}} = 40\textrm{ fm} \ .
$$
This is much larger than the expected lifetime of the hadron gas
$\lambda\simeq 5-10$ fm. He, Fries and Rapp have reported relaxation times as short as 25-40 fm in the 150-180 MeV region, in good agreement with our estimate. It should be taken into account that the hadron gas cools down to lower temperatures in the last stages, with longer relaxation times. 
Thus, charm quarks will not completely relax during the lifetime of the pion gas and will indeed carry information from the crossover from the quark-gluon plasma phase.

\begin{figure}
\centering
\includegraphics[width=8.5cm]{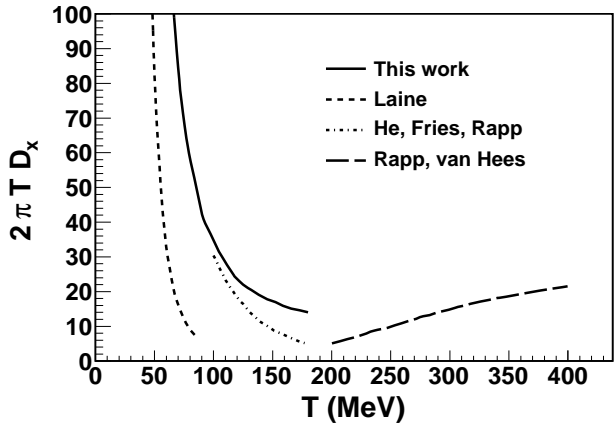}
\caption{\label{fig:comparative} 
We compare our computation of the diffusion coefficient with other estimates. The possibility of a minimum of the charm relaxation time at the phase transition seems to be well possible. The leftmost dashed line is pure perturbation theory~\cite{Laine:2011is}, to which our result seems to tend asymptotically at low temperature.
In addition, we have plotted two curves below and above the crossover~\cite{He:2011yi,Rapp:2008qc}
}
\end{figure}

In Fig.~\ref{fig:comparative} we show a comparison of our computation with the best existing ones and with the estimate of~\cite{vanHees:2005wb}
that employs a mixed plasma plus resonance approach above the crossover.
All existing information points out to that the minimum relaxation time of the charm quark happens around the phase crossover, where the interactions also have longest range and intensity. Thus charm quarks can be potentially used as a probe of the phase transition if theoretical uncertainties on the hadron gas side can be reduced. We believe that we have produced a very reliable estimate of the hadron coefficients in the temperature region $T\le 150$ MeV\footnote{Our numerical data for scattering amplitudes or transport coefficients is at the disposal of interested colleagues who want to pursue kinetic or hydrodynamic simulations by contacting {\tt{fllanes@fis.ucm.es}} or {\tt{j.torres@fis.ucm.es}}.}

Laine finds a formula for the momentum space diffusion coefficient 
$$
\Gamma_0 \propto \frac{T^7}{f_\pi^4}
$$
provided that $m_\pi/\pi \ll T \ll f_\pi$, which is a very restrictive range of temperatures around 60 MeV.
We have shown that this growth with temperature is way too fast and that properly unitarizing the interaction tames this high power of the temperature. We also qualify the statement that, in this range, the coefficients are dependent only on the pion mass and decay constant; this should be understood as valid in the infinite quark mass limit, while in the charm sector we find that the $m_D^*-m_D$ mass difference brings about non-negligible corrections.

Svetitsky and Uziel (Fig.~1 in \cite{Svetitsky:1996nj}) found that a $c$-quark with initial transverse momentum 2 GeV would have come down to 1 GeV by the time of freeze out. What our results show is that all this decrease needs to be assigned to the quark and gluon plasma phase and, especially, to the phase transition, but that the loss of momentum in the pion gas is a moderate-sized correction. For example, a 1 GeV charm quark entering the pion gas at 150 MeV and travelling four femtometers through it, will have lost about 200 MeV at freeze-out.

We have found that the $F$ drag coefficient and the $\Gamma_0$ diffusion coefficient depend only mildly on the charm-quark momentum, implying that the nuclear suppression factor for charm in the pion gas can be reasonably modelled. On the contrary, we find that the $\Gamma_1$ diffusion coefficient strongly depends on momentum, so that anisotropic observables such as the elliptic flow will have a more involved dependence with momentum.

Moreover we have shown that the thermal relaxation time at 150 MeV is about 40 fm, implying that the charm quarks do carry memory of the phase transition upon exiting the hadron gas. Our results also suggest that the spatial diffusion coefficient is likely to have a minimum at the crossover to the quark and gluon plasma. 

\vspace{3.4cm}

\emph{Work supported by grants FPA 2008-00592, FIS2008-01323,
FPA2007-29115-E, FIS2006-03438, PR34-1856-BSCH, UCM-BSCH, GR58/08 910309, PR34/07-15875 (Spain)
and by the EU Integrated Infrastructure Initiative Hadron Physics Project under Grant Agreement n.227431.
The authors thank Li Sheng Geng for updating them on the current $D\pi$ meson effective Lagrangians.
Luciano Abreu thanks the hospitality at Univ. Complutense of Madrid where this work has been completed
and aknowledges finantial support from CAPES/Fundacion Carolina.
Daniel Cabrera acknowledges finantial support from Centro Nacional de F\'isica de Part\'iculas, Astropart\'iculas y
 Nuclear (CPAN, Consolider -Ingenio 2010).
Juan M. Torres-Rincon is recipient of an FPU scholarship from the Spanish Ministry of Education.}

\appendix
\section{Static Fokker-Planck equation and one-dimensional solution}
\label{OnedFK}
Suppose that the three scalar coefficients $F$, $\Gamma_0$, $\Gamma_1$ 
in Eq.~(\ref{defAyB}) do not depend on $\mathbf{p}$ 
(limit of momentum independence or static limit). 
Then one can speak of constant $F(p^2)=F$ (we will now show that this is simply a friction coefficient), and $\Gamma_0 (p^2)= \Gamma_1(p^2)= \Gamma $ (diffusion coefficient).

The Fokker-Planck equation reads
\be 
\frac{\pa f_c}{\pa t} = F \nabla_{p} \cdot (\mathbf{p} f_c) + \Gamma \nabla^2_p f_c
\ee
that can be compared with the standard diffusion equation for the concentration $C$ of a solute
\be \frac{\pa C}{\pa t} = - \mu \nabla \cdot (C \mathbf{F}) + D \nabla^2 C,\ee
where $\mu$ is the mobility, $D$ is the diffusion coefficient and $\mathbf{F}$ is the external force. Einstein's relation $D=\mu T$ relates the diffusion coefficient and the mobility.

For clarity let us concentrate on one dimension. In this simpler case, the equation 
\be \label{Rayleighs}
\frac{\pa f_c}{\pa t} = F \frac{\pa}{\pa p} (pf_c) + \Gamma \frac{\pa^2 f_c}{\pa p^2}
\ee
is known as Rayleigh's equation and describes the momentum distribution equation for a Brownian particle.

\begin{figure}
\centering
\includegraphics[width=8.5cm]{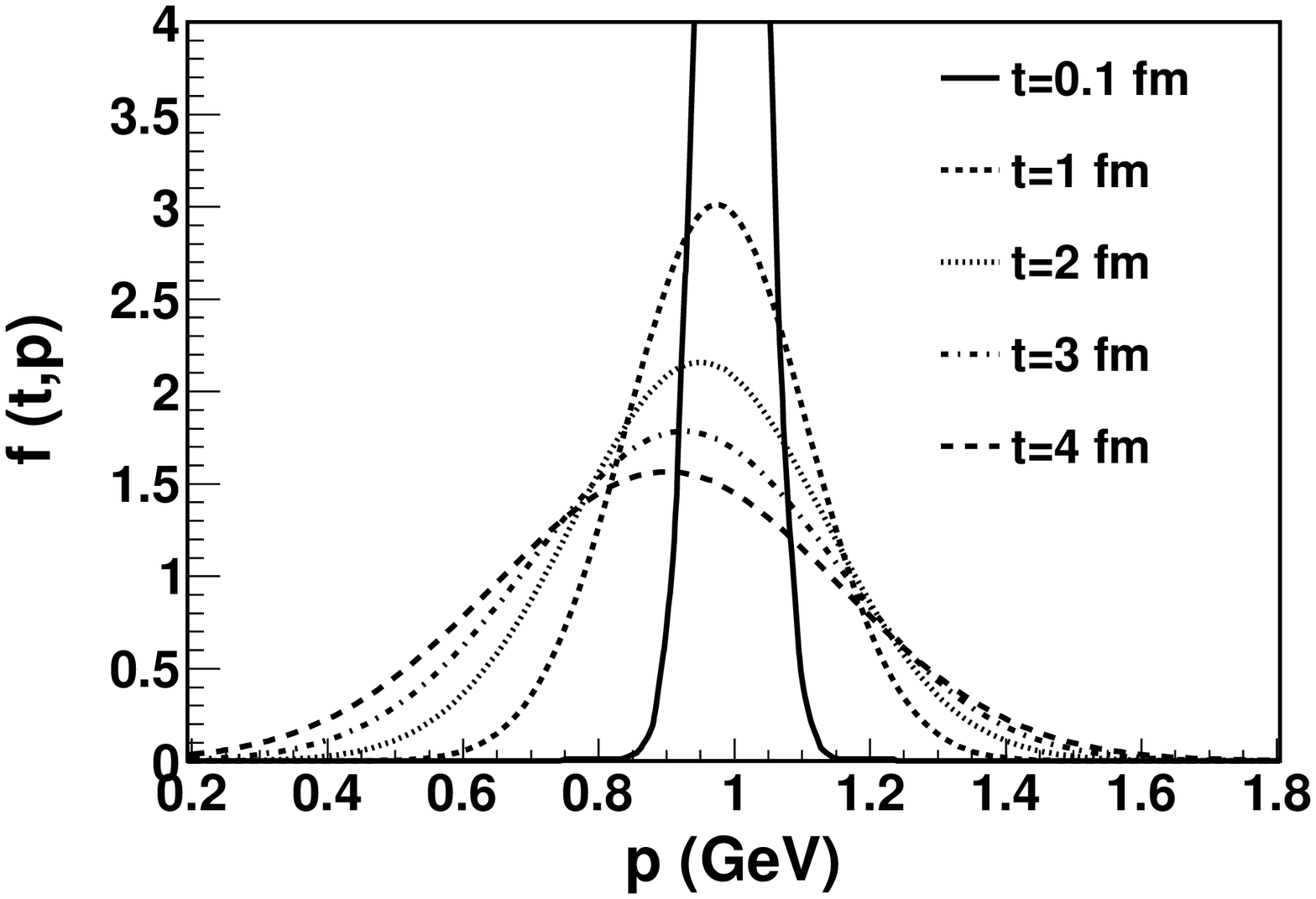}\\
\includegraphics[width=8.5cm]{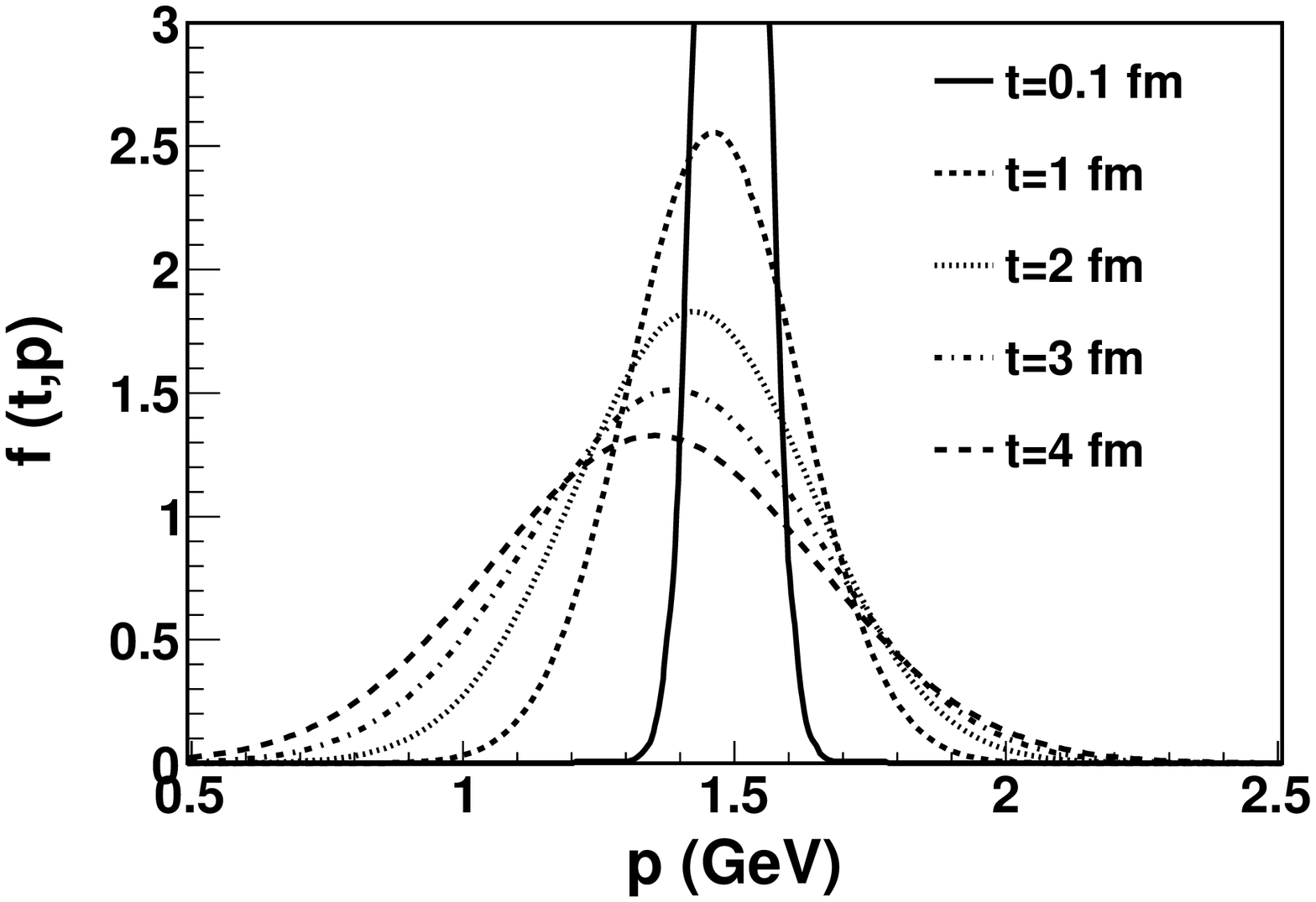}\\
\includegraphics[width=8.5cm]{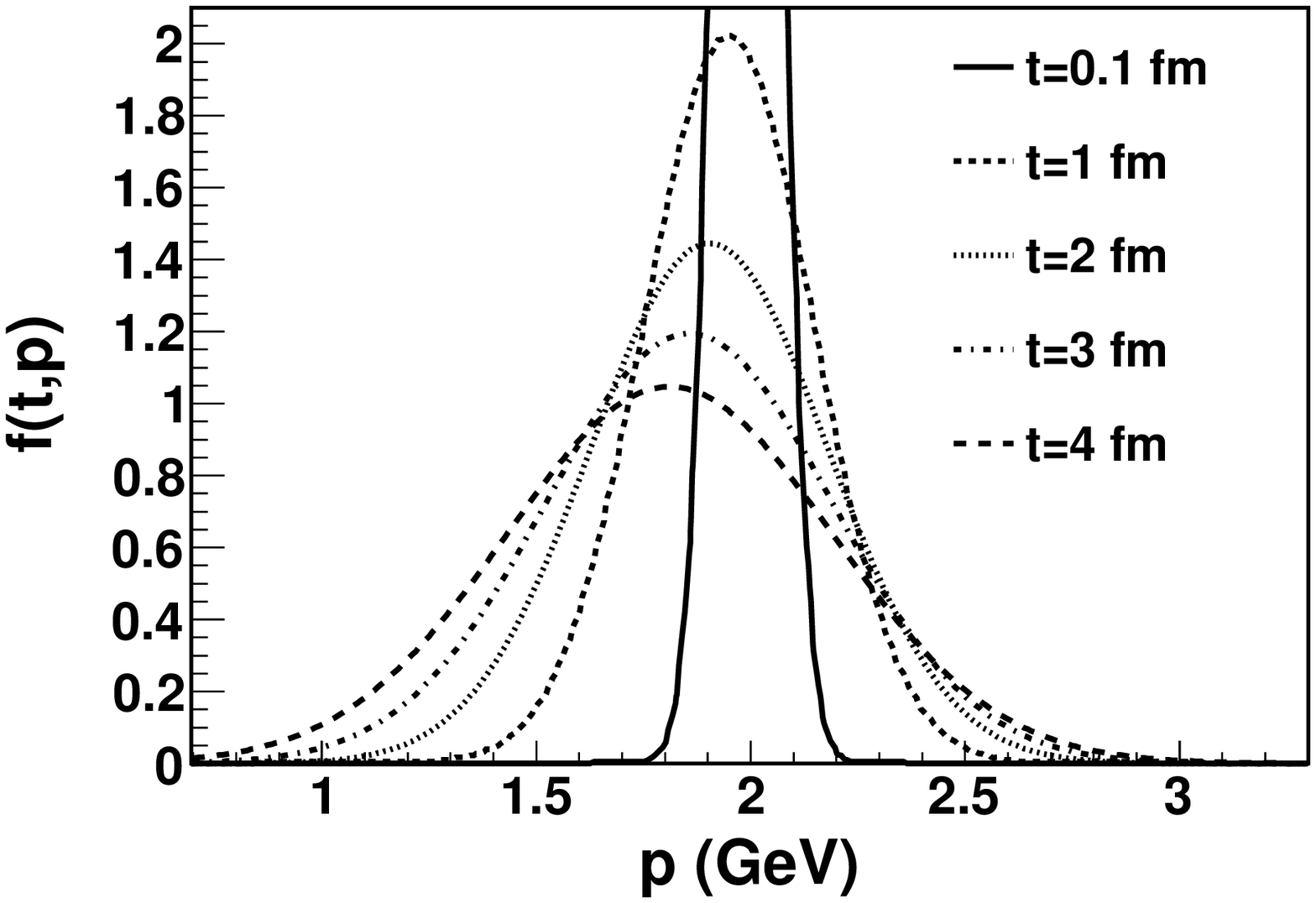}
\caption{\label{fig_1dsol} 
One-dimensional evolution of a drop of charm $f(t,p)$ from Eq. (\ref{eq:1dsol}) with well-defined momentum $p_0=1$, 1.5 and 2 GeV (top to bottom), at a reference temperature of 150 MeV. Time evolution drags the momentum towards zero  from the initial condition,
 Dirac's delta function in Eq.~(\ref{initcond}), and the shape broadens to adopt the Boltzmann equilibrium function.}
\end{figure}

With the initial condition
\be \label{initcond}
f_c (p,t=0)=\delta (p-p_0) 
\ee
the analytic solution reads

\be \label{eq:1dsol} f_c(p,t) =\left[ \frac{F}{2\pi \Gamma} \left( 1 -e^{-2 F t} \right) \right]^{-1/2} \exp \left[ - \frac{F}{2\Gamma} \frac{(p-p_0 e^{-Ft})^2}{1-e^{-2Ft}} \right] \ .\ee
This solution can be easily plotted, and we do so in Figure \ref{fig_1dsol} for 
momentum $p_0=1$, 1.5 and 2 GeV (top to bottom), at a reference temperature of 150 MeV.

We can identify the large-time behavior of this function with the Maxwell-Boltzmann 
equilibrium function
\be 
\lim_{t \rightarrow \infty}  f(t,p) =  f_{\textrm{MB}}(p), \ee
provided that an analogous to Einstein's relation holds
\be \Gamma = FmT, \ee
and the two coefficients $F$, $\Gamma$ are not independent but rather related by this fluctuation-dissipation relation (as shown below in
Eq.~(\ref{Einstein}) where we have derived this relation from the momentum-dependent fluctuation dissipation theorem independently of the number of spatial dimensions).
Morever, in Appendix \ref{langevin} we show that the momentum diffusion coefficient, $\Gamma$, is related to the spatial diffusion coefficient, $D_x$, as:
\be 
D_x = \frac{\Gamma}{m^2 F^2}=\frac{T^2}{\Gamma}. 
\ee

\section{\label{langevin} Classical Langevin equation for charm diffusion}

The purpose of this appendix is to show the relation
between the diffusion coefficient in space $D_x$ (that appears in Fick's diffusion Law) and the
momentum diffusion coefficient $D$, that we have estimated through the Fokker-Planck equation. This discussion is well-known from classical statistical physics, but it is enlightening to review it and makes the article self-contained. Since we use this material mostly to give the various quantities a physical interpretation, we believe that it is sufficient to limit ourselves to a purely classical discussion (as appropriate for a dilute gas).

We begin by rederiving the Fokker-Planck equation from the Langevin equations.
Several manipulations of Dirac delta distributions are easier to follow discretizing the time variable, to avoid resource to somewhat advanced functional analysis, and we will thereafter take again the continuum limit $\delta t\to 0$.
Then the classical solution to the Langevin will allow us to identify the space-diffusion term and relate it to the Fokker Planck coefficient of diffusion in momentum space.

The charm quark (Brownian particle) moves in the pion gas and it is diffused because of  collisions with these mesons.
The position and momentum of the charm quark can be regarded as stochastic variables depending on time. The classical, non-relativistic
stochastic differential equations that govern their motion are:
\ba \frac{dx^i}{dt} & = & \frac{p^i}{m_D} \\
\frac{dp^i}{dt} &= & -F^i(\mathbf{p}) + \xi^i(t) \ , \ea
where the index $i=1,2,3$ labels the space component of $\mathbf{x}$ and $\mathbf{p}$.
This equation is called the Langevin equation. The $F^i (\mathbf{p})$ is a deterministic 
drag force which depends on momentum through the collision processes and $\xi(t)$ is a stochastic term called white noise.
It verifies
\ba \label{stoc1} \langle \xi^i(t)\rangle & = & 0 \ , \\
 \label{stoc2} \langle \xi^i (t) \xi^j (t') \rangle & = & \Gamma^{ij} (\mathbf{p}) \delta(t-t') \ .\ea
In an isotropic gas one has $\Gamma^{ij} (\mathbf{p}) = \Gamma (\mathbf{p}) \delta^{ij}$.

We now discretize the time variable
\be 
t_n \equiv n \delta t;\quad  \mathbf{x}_n \equiv \mathbf{x}(t_n); \quad \mathbf{p}_n \equiv \mathbf{p}(t_n); \quad n=0,1,2,... 
\ee
and choose a mid-point discretization for ${\bf F}$~\cite{risken89a} 
\be 
F^i_n (\mathbf{p}) = F^i [  \frac{\mathbf{p}_n + \mathbf{p}_{n+1}}{2} ] \ . 
\ee
The discretized Langevin equation reads then
\ba 
{\bf x}_{n+1} & = & {\bf x}_n + \frac{{\bf p}_{n}}{m_D} \delta t \ ,\\
{\bf p}_{n+1} &= & {\bf p}_n - {\bf F}_n \delta t +  {\bf L}_n \delta t \ ,\ea

with a time average over the random noise
\be L^i_n = \frac{1}{\delta t} \int_{t_n}^{t_{n+1}} dt \ \xi^i(t) \ . \ee
From (\ref{stoc1}) and (\ref{stoc2}), $L^i_n$ verifies: 
\be \langle L^i_n \rangle =0 \ .\ee
\be \label{variance} \langle L^i_n L^j_{n'} \rangle= \frac{\Gamma}{\delta t} \delta^{ij} \delta_{nn'} \ .
\ee
(With somewhat more work one can show that the variable $L^i_n \sim \mathcal{O} (\delta t^{-1/2})$ ).

The average $\langle \rangle$ is taken with respect to the probability associated with the stochastic process. Since the stochastic variables are positions and momenta, this 
probability is nothing but the one-particle classical distribution function, $f(t,\mathbf{x},\mathbf{p})$.
Averages are then computed by means of
\be 
\langle T(t)\rangle_{X,P} \equiv \int d\mathbf{x} d\mathbf{p} T(t,\mathbf{x}_n,\mathbf{p}_n) f(t,\mathbf{x}_n,\mathbf{p}_n) \ ,
\ee
where $T(t,\mathbf{x}_n,\mathbf{p}_n)$ is any function of the stochastic variables and time.

In the Fokker-Planck equation we look for the time evolution of the distribution function itself, so we need to calculate 
the probability that a particle at time $t_{n+1}$ is at $\mathbf{x},\mathbf{p}$
\be \label{probforw} f(t_{n+1},\mathbf{x},\mathbf{p})= \langle \delta^{(3)}(\mathbf{x}_{n+1}-\mathbf{x}) \delta^{(3)}(\mathbf{p}_{n+1}-\mathbf{p})\rangle \ ,
\ee
from the distribution function at a prior time.

We introduce the discretized Langevin equation inside the deltas in (\ref{probforw}):
\ba 
\delta({\bf x}_{n+1}-{\bf x}) & = & \delta({\bf x}_n  -{\bf x} + \frac{{\bf p}_n}{m_D} \delta t) \ , \\
 \delta({\bf p}_{n+1}-{\bf p}) & = & \delta({\bf p}_n -{\bf p} + \left[ {\bf F}_n + {\bf L}_n \right] \delta t ) \ . 
\ea

Expanding the deltas up to $\mathcal{O} (\delta t)$,
\ba 
\delta(x^i_{n+1}-x^i)  = 
\nonumber \\ 
\delta(x^i_n -x^i) +\sum_j \frac{\pa}{\pa x^j_n} \delta(x^i_n-x^i) \ \frac{p^j_n}{m_D} \delta t \ , 
\\ \nonumber 
\delta(p^i_{n+1}-p^i)  =  
\\ \nonumber
\delta(p^i_n -p^i) +\sum_j \frac{\pa}{\pa p^j_n} \delta(p^i_n-p^i)  \ \left[ F^j (p_n) + L^j_n \right] \delta t \\
  +  \frac{1}{2}\sum_{j} \sum_k \frac{\pa^2}{\pa p^j_n \pa p^k_n} \delta(p^i_n-p^i) L^j_n L^k_n  \ (\delta t)^2 \ , 
\ea
and introducing these expansions inside equation (\ref{probforw}), we see that
\ba 
\nonumber f(t_{n+1},\mathbf{x},\mathbf{p})  =   \langle \delta^{(3)}(\mathbf{x}_n -\mathbf{x}) \delta^{(3)}(\mathbf{p}_n -\mathbf{p})  \rangle 
\\ \nonumber
 +  \langle \sum_j \frac{\pa}{\pa x^j_n} \delta^{(3)}(\mathbf{x}_n-\mathbf{x}) \ p^j_n \ \delta^{(3)}(\mathbf{p}_n -\mathbf{p}) \rangle \frac{\delta t}{m_D} 
\\ \nonumber  +  \langle \delta^{(3)}(\mathbf{x}_n-\mathbf{x}) \sum_j \frac{\pa}{\pa p^j_n} \delta^{(3)}(\mathbf{p}_n-\mathbf{p})  \ \left[L^j_n- F^j (p_n)  \right]   \rangle \delta t \\
  +  \frac{1}{2} \langle \delta^{(3)}(\mathbf{x}_n-\mathbf{x}) \sum_{j} \sum_k \frac{\pa^2}{\pa p^j_n \pa p^k_n} \delta^{(3)}(\mathbf{p}_n-\mathbf{p}) L^j_n L^k_n  \rangle  (\delta t)^2 \ .
\nonumber \\ 
\ea

In order to obtain $f(t_n, \mathbf{x}, \mathbf{p})$ in the left-hand side, we introduce the following identity
\ba 
\delta^{(3)}(\mathbf{x}_n-\mathbf{x}) \delta^{(3)}(\mathbf{p}_n-\mathbf{p})= 
\\ \nonumber
\int d\mathbf{z} d\mathbf{q}  \delta^{(3)}(\mathbf{x}_n-\mathbf{z}) 
\delta^{(3)} (\mathbf{z}-\mathbf{x}) \delta^{(3)}(\mathbf{p}_n-\mathbf{q}) \delta^{3} (\mathbf{q}-\mathbf{p})
\ea
and replace  the definition in Eq.~(\ref{probforw}) 
\be 
\langle \delta^{(3)} (\mathbf{x}_n -\mathbf{z}) \delta^{(3)}(\mathbf{p}_n-\mathbf{q}) \rangle = f(t_n,\mathbf{z},\mathbf{q})\ .
\ee

One obtains
\ba 
 \\ \nonumber
f(t_{n+1},\mathbf{x},\mathbf{p})  =   
\int d\mathbf{z} d\mathbf{q} \ \delta^{(3)} (\mathbf{z}-\mathbf{x}) \delta^{(3)} (\mathbf{q}-\mathbf{p}) \ f(t_n,\mathbf{z},\mathbf{q}) +
\\ \nonumber
  \int d\mathbf{z} d\mathbf{q} \  \delta^{(3)}(\mathbf{q}-\mathbf{p}) \sum_i \frac{\pa}{\pa z^i}  \delta^{(3)}(\mathbf{z}-\mathbf{x}) q^i \ f(t_n,\mathbf{z},\mathbf{q}) \frac{\delta t}{m_D} -
\\ \nonumber 
  \int d\mathbf{z} d\mathbf{q} \ \delta^{(3)} (\mathbf{z}-\mathbf{x}) \sum_i \frac{\pa}{\pa q^i}  \delta^{(3)} (\mathbf{q}-\mathbf{p})  F^i (\mathbf{q}) \ f(t_n,\mathbf{z},\mathbf{q})  \delta t +
\\ \nonumber  
 \int d\mathbf{z} d\mathbf{q} \ \delta^{(3)} (\mathbf{z}-\mathbf{x}) \sum_{ij} \frac{\pa^2}{\pa q^i \pa q^j}  \delta^{(3)} (\mathbf{q}-\mathbf{p})  \frac{\Gamma^{ij}(\mathbf{q})}{2} f(t_n,\mathbf{z},\mathbf{q})  \delta t 
\ea
where the average operation has been factorized because $p^i_n$ only depend on $L^i_{n'}$ with $n'<n$.

Now integrate by parts and finally, over $\mathbf{z}$ and $\mathbf{q}$:

\begin{widetext}
\ba \nonumber f(\mathbf{x},\mathbf{p}, t_{n+1} ) & = &  f(t_n,\mathbf{x},\mathbf{p}) - \frac{\mathbf{p}}{m_D}  \cdot \frac{\pa}{\pa \mathbf{x}} f(t_n,\mathbf{x},\mathbf{p}) \delta t 
 +  \sum_i \frac{\pa}{\pa p^i} F^i (\mathbf{p}) f(t_n,\mathbf{x},\mathbf{p}) \delta t 
 + \frac{1}{2} \sum_{ij} \frac{\pa^2}{\pa p^i \pa p^j} \Gamma^{ij} (\mathbf{p}) f(t_n,\mathbf{x},\mathbf{p}) \delta t \ . \ea
\end{widetext}

Now we can return to the continuum limit $\delta t \rightarrow 0$:

\begin{widetext}
\be \frac{\pa f(t,\mathbf{x},\mathbf{p}) }{\pa t} + \frac{\mathbf{p}}{m_D}  \frac{\pa}{\pa \mathbf{x}} f(t,\mathbf{x},\mathbf{p}) = 
\sum_i \frac{\pa}{\pa p^i} F^i (\mathbf{p}) f(t,\mathbf{x},\mathbf{p}) + \frac{1}{2} \sum_{ij} \frac{\pa^2}{\pa p^i \pa p^j} \Gamma^{ij} (\mathbf{p}) f(t,\mathbf{x},\mathbf{p}) \ . \ee
\end{widetext}

Taking the average in space 
\be \frac{\pa f_c(t,\mathbf{p})}{\pa t} = - \frac{\pa}{\pa p^i} \left[ F^i(\mathbf{p}) f_c(t,\mathbf{p})\right] + \frac{1}{2} \frac{\pa^2}{\pa p^i \pa p^j} \Gamma_{ij} (\mathbf{p}) f_c(t,\mathbf{p}) \ ,
\ee
that coincides with the Fokker-Planck equation in Eq.~(\ref{FKPL}).
Now we see that the diffusion coefficients $\Gamma_0$, $\Gamma_1$, stem from the random force in the Langevin equations, and the drag coefficient from the 
deterministic friction force there. 

In the static limit $\mathbf{p} \rightarrow 0$, we can solve the
Langevin (or, in this limit, also Uhlenbeck-Orstein) equation
\be 
\frac{d \mathbf{p}}{dt} = -F\mathbf{p} + \mathbf{\xi} (t) \ , 
\ee
whose solution is
\be 
\label{p_sol} \mathbf{p} (t)= \mathbf{p}_0 e^{-Ft} + e^{-Ft} \int_0^t 
d\tau e^{F\tau} \mathbf{\xi}(\tau) \ .
\ee
Taking the average one can see that due to the drag force, the friction term makes the particle eventually stop in the fluid's rest frame.
\be \langle \mathbf{p} (t) \rangle = \mathbf{p}_0 e^{-Ft}.\ee

The second of Hamilton's equations 
\be 
\frac{d \mathbf{x}}{dt} = \frac{\mathbf{p}}{m_D}, 
\ee
is then solved by
\be 
\label{x_sol} \mathbf{x}(t)=\mathbf{x}_0 + \int_0^t d\tau \frac{ \mathbf{p} (\tau)}{m_D} 
\ee
Or, on average,
\be 
\langle \mathbf{x} (t) \rangle =\mathbf{x}_0 + \frac{ \mathbf{p}_0}{F m_D} 
(1-e^{-Ft}).  
\ee

To make the connection with the spatial diffussion coefficient we can show the mean quadratic displacement of the Brownian particle
($r=\sqrt{x^2+y^2+z^2}$)
\be 
\langle (r(t)-r_0)^2 \rangle=\langle (x(t)-x_0)^2+(y(t)-y_0)^2+(z(t)-z_0)^2 \rangle \ ,
\ee
that,  from Fick's diffusion law, is simply 
\be 
\langle (r(t)-r_0)^2 \rangle = 6 D_x t \ . 
\ee

From the averaged solution to Langevin's equation~(\ref{x_sol}),
\be 
\langle (x(t)-x_0)^2 \rangle = \frac{1}{m_D^2} \int_0^t \int_0^t d\tau d\tau' \ \langle p^x (\tau) p^x(\tau') \rangle \ . 
\ee

With the help of (\ref{p_sol}) and (\ref{stoc2}) and carefully performing the integral \cite{risken89a} one obtains the leading term of this 
expression when $t \gg F^{-1}$ as
\be 
\langle (x(t)-x_0)^2 \rangle = \frac{2 \Gamma t}{m^2_D F^2} \ ,
\ee
so that
\be 
D_x = \frac{\Gamma}{m^2_D F^2} = \frac{T^2}{\Gamma} \ ,
\ee
where finally we have used Einstein's relation. Thus, the calculation of the momentum space diffusion coefficient automatically entails an estimate for the space diffusion coefficient.

\section{Fluctuation-Dissipation Relations}\label{appendixFDR}

Not all three coefficients $F(p^2)$, $\Gamma_0 (p^2)$ and $\Gamma_1 (p^2)$ 
appearing in the Fokker-Planck equation
are  independent, but rather
related by a  fluctuation-dissipation relation.
This just means that thermal equilibrium requires the damping force $F$ to match the fluctuations of the charm quark momentum distribution so as to maintain energy equipartition, with $\frac{k_B T}{2}$ per degree of freedom. Since we consider the $p$-dependence of the three coefficients, the fluctuation-dissipation relation will be momentum dependent, although we also expose the $p\to 0$ limit. More details on deriving such relations can be found in standard textbooks~\cite{landau1981course}.

A transparent procedure is to  match the asymptotic solution of the Fokker-Planck equation to the thermal equilibrium distribution function, thus guaranteeing energy equipartition.

First of all, the Fokker-Planck equation can be written as an equation of continuity
\be \frac{\pa f_c (t, \mathbf{p})}{\pa t} =- \frac{\pa}{\pa p_i} n_i \ , \ee
where
\be n_i \equiv -F_i (p^2) f_c (t,\mathbf{p}) - \frac{\pa}{\pa p_j} \left[ \Gamma_{ij} (p^2) f_c (t,\mathbf{p}) \right]\ee
is the particle flux density in momentum space.
At statistical equilibrium, this flux is zero, and the equilibrium distribution function is the Bose-Einstein function,
\be f_c \sim \frac{1}{e^{-p^2/2 M T}-1} \ . \ee

Employing again the approximation $1 + f_c \approx 1$ valid for small charm-quark number, one can obtain
\be 
F_i (p^2) + \frac{\pa \Gamma_{ij} (p^2)}{\pa p_j} = \frac{1}{MT} \Gamma_{ij}(p^2) p_j \ .
\ee
This momentum-dependent fluctuation-dissipation relation can be recast for the functions $F(p^2), \Gamma_0 (p^2)$ and $\Gamma_1 (p^2)$ as:
\be F (p^2) + \frac{1}{p} \frac{\pa \Gamma_1 (p^2)}{\pa p} + \frac{2}{p^2} \left[ \Gamma_1 (p^2)
- \Gamma_0 (p^2) \right] =  \frac{\Gamma_1 (p^2)}{MT} \ . \ee

\begin{figure}
\centering
\includegraphics[width=8.5cm]{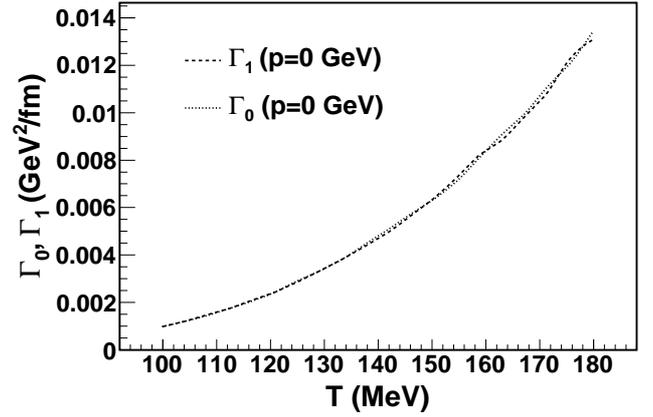}
\caption{\label{fig:momdiff} 
Momentum-space transport coefficients as function of temperature.
Dotted: $\Gamma_0 (p^2 \to 0)$. Dashed: $\Gamma_1 (p^2 \to 0)$.
The very good agreement in our computer programme, as appropriate in this limit, makes the curves barely distinguishable.
}
\end{figure}

\begin{figure}
\centering
\includegraphics[width=8.5cm]{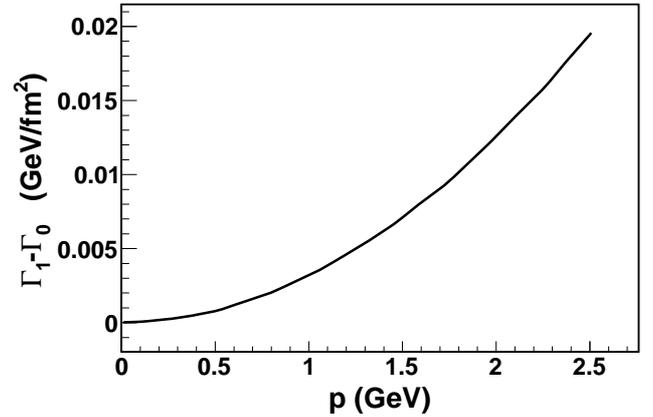}
\caption{\label{fig:momdiff2} 
Momentum-space transport coefficients as function of momentum
at fixed temperature 150 MeV. The two coefficients converge at low $p$.
}
\end{figure}

For low-momentum charm quarks, $\Gamma_1(p^2), \Gamma_0 (p^2) \rightarrow \Gamma$, $F(p^2) \rightarrow F$. The equality of the two $\Gamma$ coefficients in the limit of zero momentum is numerically checked in
 Figs.~\ref{fig:momdiff} and \ref{fig:momdiff2}.

We then recover the
well known Einstein relationship
\be 
\label{Einstein} F = \frac{\Gamma}{MT} 
\ee
which is the same result derived above in Appendix~\ref{OnedFK} for the one-dimensional solution of the Rayleigh equation.

Thus, in the static limit two coefficients take the same value
and the third is obtained from them by Eq.~(\ref{Einstein}), and
we are left with only one independent diffusion coefficient.

The Langevin equation also allows us to directly obtain the classical interpretation of $F$ as a loss of energy per unit length. Ignoring the fluctuating force,
\be
\frac{d\gamma m{\bf v}}{dt} = -{\bf F}
\ee
can be multiplied by ${\bf v}$ to yield after some reshuffling the obvious expression for the power
\be
\frac{dm\gamma}{dt} = -{\bf F}\cd {\bf v} 
\ee
and remembering the definition ${\bf F}=F{\bf p}$ in Eq.~(\ref{defAyB}), the loss of energy per unit length is simply $F\ar{\bf p}\ar$, as in the non-relativistic theory.\\
The loss of momentum per unit length can then be expressed as
\be
\frac{d{\bf p}}{dx} = 
\frac{d{\bf p}}{vdt} = - F E
\ee
in terms of the energy and momentum of the charmed particle.

\section{Kinematics}\label{Kinematicsapp}

In the evaluation of the drag and diffusion coefficients in equation~(\ref{Transportintegrals}) we need to calculate integrals of the generic type
\ba \label{integraltocalc} \nonumber
g_{\pi}\! \int  d\mathbf{k} \frac{d\mathbf{q}}{(2\pi)^9} 
 f_{\pi} (\mathbf{q}) \left[ 1+f_{\pi} (\mathbf{q} + \mathbf{k})\right]
 \frac{1}{2E_q^{\pi}} \frac{1}{2E_p^c} \frac{1}{2E_{q+k}^{\pi}} 
\frac{1}{2E^c_{p-k}} 
\\ \nonumber  
(2\pi)^4 \delta (E_p^c + E_q^{\pi}-E^c_{p-k} -E^{\pi}_{q+k}) 
\! \sum |\mathcal{M}_{\pi c}^2 (s,t,\chi)|^2 \! \mathcal{G} (k^i,p^i) \\
\ea
where the various cases differ in the choice of $\mathcal{G}$ kinematic function
\be 
\mathcal{G} (k^i,p^i) = \left\{
\begin{array}{ccc}
\frac{k_i p^i}{p^2} \ & \textrm{ for} & F(p^2) \\
\frac{1}{4} \left[ \mathbf{k}^2 - \frac{(k_i p^i)^2}{p^2} \right] \ & \textrm{ for} & \Gamma_0(p^2) \\
\frac{1}{2} \frac{(k_i p^i)^2}{p^2} \ & \textrm{ for} & \Gamma_1(p^2)
\end{array} \right.
\ee

The collision momenta are labelled as
\be c(\mathbf{p}) + \pi (\mathbf{q}) \rightarrow c(\mathbf{p-k}) + \pi (\mathbf{q+k}) \ee
so that $P^{\mu}=(E_p^c,\mathbf{p})$ is the 4-momentum of the incoming $c$ quark, $Q^{\mu}=(E_q^{\pi}, \mathbf{q})$ the 4-momentum of the incoming pion,
and $\mathbf{k}$ the transfered momentum from the $c$ quark to the pion. 
The $c$ quark can be in a $D$ or in a $D^*$ meson states, degenerate in Leading Order Heavy Quark Effective Theory. We generically use an average $m_D$ for the transport code, although we distinguish the masses in the scattering amplitude to correctly position the $D_0$ and $D_1$ resonances.
For example, the outgoing-particle energies are $E^c_{p-k} = \sqrt{m_D^2 + (\mathbf{p-k})^2}$ and  $ E^{\pi}_{q+k}=\sqrt{m^2_{\pi} + (\mathbf{q+k})^2}$ respectively.

The resulting transport coefficients obtained after integrating 
Eq.~(\ref{integraltocalc}) depend only on the modulus of $\mathbf{p}$. However we will
introduce a (trivial) $d\Omega_p$ angular integration in the $\mathbf{p}$-coordinates
\be 
\int d \mathbf{k} \ d\mathbf{q} \rightarrow \int d\mathbf{k} \ d\mathbf{q} \ \frac{d \Omega_p}{4\pi} \ ,
\ee
in order to increase our freedom in the choice of axes.

We also find convenient to change the integration variables from the incoming pion and transfered momenta, $\mathbf{q}$ and $\mathbf{k}$ respectively, to the total momentum
$\mathbf{P} = \mathbf{p}+ \mathbf{q}$ and  the outgoing charm
momentum $\mathbf{p}_3 = \mathbf{p} - \mathbf{k}$. The Jacobian determinant associated to these translations is unity.

Now, without loss of generality, we choose the total momentum $\mathbf{P}$ vector along the $OZ$ axis, and the incoming charm momentum $\mathbf{p}$ lying on the $OZX$ plane. Automatically $\mathbf{q}$ is in this plane as well.
Finally, the outgoing charm momentum $\mathbf{p}_3$ has in general all three Cartesian projections,
\ba
\mathbf{P} = (0,0,P) \\
\mathbf{p} = (p \sqrt{1-x_p^2},0,px_p) \\
\mathbf{q} = \mathbf{P}-\mathbf{p}=(-p \sqrt{1-x^2_p},0,P-p x_p) \\
\frac{\mathbf{p}_3}{p_3} = (\sqrt{1-x^2_3} \cos \phi_3 , \sqrt{1-x^2_3} \sin \phi_3 , x_3) \ . 
\ea
Here $x_p$ is the cosine of the polar angle of $\mathbf{p}$, that is, of the relative angle between $\mathbf{p}$ and $\mathbf{P}$;
$x_3$ and $\phi_3$  the cosine of the polar angle and azimuthal angle associated with $\mathbf{p}_3$.
The transfered $\mathbf{k} = \mathbf{p}-\mathbf{p}_3 $
and outgoing pion 
$ \mathbf{p}^{\pi}_4 = \mathbf{q} + \mathbf{k} = \mathbf{P} - \mathbf{p}_3 $
momenta are then dependent variables.

The angular integrals associated with $\mathbf{P}$ are then trivial (the scattering matrix is rotation-invariance) and yield $4\pi$ (they have {\emph{de facto}} being exchanged for $d\Omega_p$ that is now non-trivial). The system has one more rotational invariance, as holding the  $\mathbf{P}$ axis fixed, one can rigidly rotate all other vectors around it~\cite{Dobado:2003wr} (so our choice of $OX$ axis does not imply any loss of generality). This trivializes the $\phi_p$ integration.

With such choice of axes the integration measure can be explicitly written down as
\ba
\int d \mathbf{k} \ d\mathbf{q} \ \frac{d \Omega_p}{4\pi} = \int d\mathbf{P} \ d\mathbf{p}_3 \ \frac{d \Omega_p}{4\pi } \\ \nonumber
= \int 4\pi  P^2  dP\ p_3^2dp_3  d\phi_3 dx_3 \ \frac{1}{4\pi} 2\pi dx_p \\ \nonumber
= 2\pi \int dP P^2 dp_3 p_3^2 d \phi_3 dx_3 dx_p
\ea

Energy conservation imposes an additional restriction, which is very non-linear in terms of the momentum variables
\ba 
E^c_p= \sqrt{m_D^2 + p^2 } \\
E^{\pi}_q= \sqrt{m_{\pi}^2 +P^2+ p^2-2 P p x_p } \\
E^c_{p-k}= \sqrt{m_D^2 + p_3^2 } \\
E^{\pi}_{q+k}= \sqrt{m_{\pi}^2 + P^2 + p_3^2 -2 P p_3 x_3} \ . 
\ea
To solve the restriction  we introduce a further auxiliary variable $W$, which is an off-shell extension of the total energy~\cite{Manuel:2004iv}, by means of
\ba
\delta (E_p^c + E_q^{\pi} - E_{p-q}^{c} -E^{\pi}_{q+k} )= \\ \nonumber
\int dW \delta (E_p^c + E^{\pi}_q -W) \delta (W-E_{p-q}^c-E^{\pi}_{q+k}) \ .
\ea
The square roots are now easier to handle two at a time, and the Dirac delta functions can be used to eliminate the two polar cosines, leaving behind only an integration over the auxiliary $W$ variable and no deltas.

The first delta 
$$\delta(E_p^c + E_q^{\pi} -W) = \frac{E_q^{\pi}}{Pp} \delta(x_p -x_{p_0})$$ 
can be used to integrate over $x_{p}$ and fix it to
\be 
x_{p_0} =\frac{P^2 + m^2_{\pi} - m^2_D + W (2 E_p^c -W)}{2Pp} 
\ee
and likewise, the second one 
$$  \delta(W-E_{p-q}^c - E_{q+k}^{\pi}) = 
\frac{E_{q+k}^{\pi}}{Pp_3} \delta(x_3 -x_{3_0})$$  
provides the $x_{3}$ integration and fixes the variable to
\be 
x_{3_0} =\frac{P^2 + m^2_{\pi} - m^2_D + W (2 E_{p-q}^c -W)}{2Pp_3} \ . 
\ee

With this kinematic work, the integrals in Eq.~(\ref{integraltocalc}) have been reduced to a four-dimensional integration 
\ba
\frac{g_{\pi}}{256 \pi^4} \int dP dW dp_3 d\phi_3 \frac{p_3}{p E_p^c E_{p-q}^c} 
\\ \nonumber
\times f_{\pi} (\mathbf{P}-\mathbf{p}) \left[ 1 + f_{\pi} (\mathbf{P}-\mathbf{p}_3)\right]
\\ \nonumber
\times \sum | \mathcal{M}(s,t,\chi)|^2 \ \mathcal{G} (k^i,p^i)\ .
\ea
We employ standard Montecarlo methods in a computer programme to numerically calculate these integrals. In particular we employ the well-known VEGAS algorithm as coded by P.~Lepage~\cite{Lepage:1980dq}. The ultraviolet integration is cutoff by the Bose-Einstein factors, and since we retain the pion masses there are no infrared enhancements. Convergence is rapidly achieved.


\end{document}